\documentclass[fleqn,10pt]{SelfArx} 
\usepackage[english]{babel} 
\usepackage{lipsum} 
\usepackage{graphicx}

\definecolor{color1}{RGB}{0,0,90} 
\definecolor{color2}{RGB}{0,20,20} 

\usepackage{hyperref} 

\JournalInfo{ArXiv} 
\Archive{} 

\PaperTitle{A Psychologically-Motivated Model of Opinion Change with Applications to American Politics} 

\Authors{Peter Duggins} 
\affiliation{Computational Neuroscience Research Group, University of Waterloo} 
\affiliation{psipeter@gmail.com} 

\newcommand{\keywordname}{Keywords} 

\Abstract{Agent-based models are versatile tools for studying how societal opinion change, including political polarization and cultural diffusion, emerges from individual behavior. This study expands agents' psychological realism using empirically-motivated rules governing interpersonal influence, commitment to previous beliefs, and conformity in social contexts. Computational experiments establish that these extensions produce three novel results: (a) sustained ``strong'' diversity of opinions across society, (b) opinion subcultures, and (c) pluralistic ignorance. These phenomena arise from a combination of agents' intolerance, susceptibility and conformity, with extremist agents and social networks playing important roles. The distribution and dynamics of simulated opinions reproduce two empirical datasets on Americans' political opinions.}

\begin{document}
\flushbottom 
\maketitle 
\tableofcontents 
\thispagestyle{empty} 


\section*{Introduction}
\addcontentsline{toc}{section}{Introduction}

Opinions are mutable: individuals revise their beliefs through social interaction, personal experience, and reflection, while societal norms shift in response to global events and public opinion. Opinion change at the individual and societal scales interact to produce political polarization, cultural globalization, and other important social trends. To understand these phenomenon and design appropriate interventions, we need quantitative tools that simulate the psychological and social aspects of opinion change. For example, models of interpersonal communication will help activists organize grassroots support, help leaders design effective campaigns, and help peacekeepers prevent the spread of extremism. Computational models of opinion change have studied the relationship between polarization, social influence, and political intolerance \cite{latane1981,hegselmann2002,deffuant2002,jager2005,baldassarri2007}, while models of cultural diffusion have improved our understanding of cultural convergence \cite{carley1991,mark1998}, subculture formation \cite{bednar2010,kitts2006}, and cultural stability within organizations \cite{carroll2002,kitts2007}. 

Building multi-level, quantitative, predictive models of opinion change is challenging because opinions arise from a multitude of neurological, psychological, and social processes. Empirically, the extent to which people are persuaded by each others' subjective evaluations depends on numerous factors, including previous beliefs and a desire to minimize cognitive dissonance \cite{petty1997}; motivations to be accurate, self-consistent, and socially accepted \cite{wood2000,cialdini2004}; issue framing, emotional arousal, and cognitive elaboration \cite{gawronski2006}; self-esteem \cite{pool1998}; social norms \cite{lapinski2005}; and more. Mathematical and computational models help formally investigate both the interplay of internal psychological forces and the feedback between opinion change and social influence among many individuals. Unfortunately, models have historically neglected important elements of social psychology, assuming that individuals behave identically, rationally, or with perfect information. This raises questions about whether their results properly inform our understanding of human societies.

Agent-based models (ABMs) seek to explain macroscopic outcomes by showing that artificial societies populated by psychologically-plausible software individuals can, when initialized in a virtual environment and evolved through time, endogenously ``grow'' complex social phenomenon \cite{epstein2006}. Three features of ABMs make them ideal for modeling opinion change. First, agents are autonomous and heterogeneous: each individual has distinct internal attributes, such as an intolerance of opposing views, a propensity to socially conform, or a tendency towards stubbornness. Second, agents can be psychologically and cognitively authentic, endowed with rational, emotional, and social thinking of arbitrary complexity \cite{epstein2014}. Third, agents interact locally in an explicitly defined space: individuals have incomplete information about the world, and interact in social networks of plausible size and composition, causing influence to spread through society in a manner constrained by personal connections. 

Although a rich literature of opinion dynamics using ABMs already exists \cite{latane1981,degroot1974,hegselmann2002,deffuant2002,dandekar2013,salzarulo2006,jager2005,axelrod1997,carley1991,mark1998}, several important questions remain unanswered:

\begin{enumerate}
\item \textit{How do social groups maintain a diversity of opinions?} Previous models have shown that when agents exchange interpersonal influence, their opinions either converge to a single value (consensus) or diverge to homogeneous opinion groups (polarization). Although consensus and polarization are important political and cultural trends, real societies never converge or diverge absolutely: diversity is always preserved. Surprisingly, models have not yet shown that such a distribution can persist. 
\item \textit{Will subcultures of opinions survive in a well-connected society?} Pockets of extreme opinions exist within moderate real-world societies. Although such subcultures have emerged in previous models, they survive only because of psychologically-implausible rules that curtail any interpersonal influence.
\item \textit{Does pluralistic ignorance affect societal opinion change?} The views we express in public often differ from those we hold privately. This situation undoubtedly affects individual and societal opinion dynamics, yet falsification remains unstudied in computational models.
\item \textit{Can an opinion change model reproduce empirical data?} Opinion dynamics models capture qualitative phenomenon like polarization and clustering, but are rarely validated with quantitative empirical data. Closing the loop will increase the scientific credibility and predictive power of these models.
\end{enumerate}

In this study, I aim to answer these questions by studying the relationship between the social psychology of personal opinion change and the distributions, dynamics, and geography of opinions across society. In Section 2, I review the literature on the social and psychological forces that drive opinion change. In Section 3, I describe the model, explaining how it extends previous models by expanding the psychological realism of agents. In Section 4, I pose hypotheses about the relationship between psychological forces and societal opinion change, run computational experiments to test them, and describe the emergence of (a) strong societal diversity, (b) persistent subcultures of opinions, and (c) pluralistic ignorance. In Section 5, I compare these results with empirical data on Americans' political opinions. I conclude by summarizing the major findings, suggesting extensions to the model, and proposing a research agenda for agent models in the social sciences.


\section*{Social Psychology of Opinion Change}
\addcontentsline{toc}{section}{Social Psychology of Opinion Change}

\textbf{Social influence} is a process in which the social exchange of information causes individuals to reevaluate their own opinions on a subjective issue. Arguably the most important feature of social influence is homophily, the principle that contact between similar people occurs more frequently and has greater impact than contact between dissimilar people. Empirical evidence for homophily and its effects on social influence abounds: for an overview, see \cite{mcpherson2001}. Interpersonal influence among friends is known to engender common attitudes \cite{friedkin1984, marsden1988, friedkin1993}, while the strength of dyadic connections concurrently increases with similarity \cite{carroll2002,kitts1999}. On the other hand, interactions can impart negative social influence if opinions differ greatly \cite{rosenbaum1986, smeaton1989}, causing individuals to adopt more extreme attitudes when exposed to counterattitudinal arguments \cite{lord1979,miller1993,taber2006}.

Homophily is a cornerstone of opinion dynamics models: individuals exert social influence on each other proportional to their ideological similarity. In dyadic conversations, similarity encourages consensus, while dissimilarity fosters polarization. A lineage of models have shown that a society with high \textbf{tolerance} (a parameter governing the relationship between opinion similarity and the magnitude of influence) leads to consensus, while low tolerance leads to polarization \cite{hegselmann2002,deffuant2002,latane1981,degroot1974,dandekar2013,salzarulo2006,jager2005,axelrod1997,carley1991,mark1998}. \textbf{Weak diversity}, defined as the convergence of opinions to $n>1$ attractor states, can be maintained when opinion subcultures form and become isolated. This outcome is common in bounded confidence models when influence between dissimilar agents goes to zero. Generally, \textbf{strong diversity}, defined as a smooth distribution of opinions along a continuous ideological spectrum, disappears in these models whenever social networks are fully connected \cite{kitts2007, flache2011, mas2014}, even accounting for noise and other minor deviations \cite{klemm2003, de2009}. 

Social influence does not take place in a vacuum, but in an environment filled by people who seek social acceptance and who judge each other upon personality and beliefs. \textbf{Conformity} describes an individual's desire to gain social approval and avoid rejection by expressing normative beliefs. There is substantial empirical evidence of people misrepresenting their true beliefs \cite{wood2000,cialdini2004,asch1951}, though some ``anticonformists'' will express non-normative beliefs so as to appear more distinct \cite{mas2014,imhoff2009}. Together, conformity and distinctiveness lead to \textbf{pluralistic ignorance} \cite{prentice1993}, a condition in which the true distribution of opinions in society differs from what is spoken and heard in public. Pluralistic ignorance makes people unaware of others' true beliefs; a lack of accurate information can, though the mechanisms of social influence, feedback to change people's true opinions. For example, after years government oppression, levels of popular dissent in authoritarian societies may become suddenly obvious, leading to political turmoil and violent tipping points \cite{kuran1989,goodwin2011}. Despite current enthusiasm for studying the effects of conformity and distinctiveness on opinion change \cite{jarman2015,mas2014,smaldino2015}, ABMs have yet to investigate the repercussions of agents' explicit belief falsification on public opinion.

The way an individual receives and internalizes others' beliefs can be as important as the content and context of the influence. People who hold strong opinions are \textbf{committed} to their beliefs: they resist opinion change, because it would challenge their political worldview and induce cognitive dissonance, and because they judge contrary information as invalid due to confirmation bias \cite{ajzen2001}. Strongly opinionated individuals have been shown to reject opinions contrary to their own belief and even become more extreme. On the other hand, moderately opinionated individuals are \textbf{susceptible} to opinion change and will more readily internalize beliefs presented by others \cite{lord1979, miller1993, taber2006}. Surprisingly, few models of opinion change have looked into how susceptibility and commitment help sustain diversity and prevent homogenization of small cultural groups \cite{baldassarri2007}.

Finally, the social networks through which individuals interact determine how opinion change spreads through society. These networks can be characterized by statistical descriptions such as the degree of connectivity (average size of a social network); real-world networks have positive assortativity (people with large networks tend to know others with large networks), low whole-network density (most people don't know each other), and high but heterogeneous clustering. Though simulations have confirmed that the size and composition of social networks strongly affect opinion change, their outcomes vary widely with the models' assumptions about the network \cite{amblard2004,centola2005}, which rarely take these empirical regularities into account. One procedure which does effectively reproduce these statistics is the social circle model \cite{hamill2009}, which is easily incorporated into an ABM framework \cite{zu2013}.


\section*{The Influence, Susceptibility, and Conformity Model (ISC)}
\addcontentsline{toc}{section}{Model Description}

To summarize, agents are randomly placed within a two-dimensional space. Each agent has a unique initial opinion, three parameters for tolerance, conformity, and susceptibility, and a social network. Each round, every agent initiates a dialogue with members of his social network. In the dialogue, each agent expresses an opinion that reflects his true opinion, his conformity, and the opinions already expressed in the dialogue. Afterwards, the initiating agent updates his true opinion based on his tolerance, susceptibility, and the expressed opinions' weighted influence. The model records the true and expressed opinion of each agent after every round. The model, data, and figures are available on \href{https://github.com/psipeter/influence_susceptibility_conformity}{GitHub}.

Agents' opinions, interpreted as beliefs on a single subjective issue, lie on a continuous $0-100$ scale. Initial opinion, tolerance, conformity, susceptibility, and social reach are all drawn from normal distributions whose means and variances are specified in each experiment. Agents are randomly assigned a continuously-valued $(x,y)$ location, then each agent creates a social network $N$ with all agents within euclidean radius equal to his social reach $r$, as per the social circle model. Agents remain stationary.

Agent $i$ initiates a dialogue with all agents $j$ in his social network. He is the first to express an opinion, and always voices his true opinion $(O_i)$. Subsequently, each $j$ distorts his opinion in order to conform or appear distinct. Specifically, $j$ calculates the average of all opinions $(E_k)$ expressed so far in the dialogue $(D)$, then expresses an opinion $(E_j)$ that is between his true opinion $(O_j)$ and the dialogue's opinion norm (conformity), or that is distanced by some amount from the dialogue's norm (distinctiveness):
\begin{equation}
E_j = O_j + \frac{c_j}{k_j} * \frac{1}{N}\sum_k^{D} (E_k - O_j)
\end{equation}
The agent parameter $c_j$ represents an agent's inherent willingness to misrepresent his beliefs in social contexts in order to appear either normal or distinct. The parameter captures both conformity $(c_j>0)$ and distinctiveness $(c_j<0)$. Greater magnitude $c_j$ produces greater belief falsification: $c_j=0$ causes the agent to speak truthfully, $c_j=1$ causes the agent to express the dialogue's ``mean opinion'', and $c=-1$ causes the agent to express an opinion that is more dislike the mean than his true opinion. In this model, conformity and distinctiveness are manifest in expression but not directly in opinion change: agents attempt to gain social favor by stating opinions that differ from their true beliefs, but do not change their true beliefs to reflect this posturing.

The extent of $j$'s conformity is further mitigated by his current commitment $k_j$, which is proportional to his susceptibility $s_j$ and the extremeness of his current opinion:
\begin{equation}\label{commitment}
k_j  = 1 + s_j * \frac{|50 - O_j|}{50}.
\end{equation}
The susceptibility parameter $s_j$ represents an agent's inherent commitment to strong beliefs; it causes him to be less affected by social context and social influence. Its magnitude governs how a departure from a neutral opinion $(O_i=50)$ translates to a shrinking of influence: higher values result in less opinion change.

After each $j$ has expressed $E_j$ once in the dialogue, $i$ updates his true opinion according to the dialogue's influence $(I_i)$, which is proportional to each $E_j$ and the weight that $i$ assigns to that expression $(w_{ij})$:
\begin{equation}
I_i = \frac{\sum_j^N w_{ij} * (E_j - O_i)}{\sum_j^N |w_{ij}|}
\end{equation}
Conceptually, the dialogue's influence $I_i$ results from $i$ being pulled towards (or pushed away from) each opinion expressed in the dialogue, $E_j$, by an amount proportional to the interagent weight, $w_{ij}$. The weight, in turn, is calculated according to homophily: the greater the absolute distance between $i$'s opinion and $j$'s expression, the more negative the weight, and the less influence $j$'s expression will exert on $i$'s opinion: 
\begin{equation}\label{weight}
w_{ij} = 1 - t_i \frac{|E_j - O_i|}{50}
\end{equation}
where $t_i$ represents $i$'s inherent intolerance of dissimilar opinions. Its magnitude dictates how strongly a given opinion difference translates to a loss of interagent weight. A high value implies that an agent will only assign positive weight to opinions that are similar to his own beliefs; a low value implies the agent will be positively influenced by a wider range of opinions. Mathematically, $t_i$ is the slope of $i$'s weight vs. $\Delta$ opinion curve, which is continuous and linear. This is a departure from the canonical bounded confidence approach, in which weight is a threshold function of an agent's intolerance $\epsilon_i$. I believe continuous weighting better reflects the subtleties of opinion appraisal and social influence than a binary ``full acceptance vs. complete disregard'' judgment. This approach has also been adopted by \cite{mas2014}. Weights are bounded from $-1$ to $+1$.

Finally, $i$ updates his true opinion based on his previous opinion and the dialogue's influence, scaled by his commitment:
\begin{equation}
O_{i,t+1} = O_{i,t} + \frac{I_i}{k_i}
\end{equation}
This process is repeated for each $i$ in the population, concluding one timestep.

I use four metrics to investigate the diversity, dynamics, and geography of opinions within the population. Opinion histograms plot the frequency of opinions across the ideological spectrum at particular times, and are the most complete measure of strong vs. weak diversity. Opinion trajectories plot each agent's history as a line on a opinion vs. time graph, and are used to study dynamics towards or away from diversity. To distinguish different regions of opinion space, I use the terms \textbf{centrist} to describe agents who hold ($33<O_i<66$), \textbf{moderate} to describe agents with moderately-strong opinions ($16<O_i<33$ or $66<O_i<83$), and \textbf{extremist} to describe agents with the strongest opinions ($0<O_i<16$ or $83<O_i<100$). Spatial maps plot each agent as a circle in $(x,y)$ space with color representing the agent's opinion, and can help identify subgroup formation and regions of ideological mixing. Finally, the \textbf{Jensen-Shannon Divergence} (JSD) is a measure of the similarity of true opinions and expressed opinions across society. The JSD quantifies pluralistic ignorance and is used to study how agents' falsifications affects the diversity of opinions within society. It is calculated from the Kullback--Leibler divergence $D(P||Q)$, a standard entropy metric for probability distributions:
\begin{equation}
JSD(P||Q)=\frac{1}{2} D(P||M)+\frac{1}{2} D(Q||M)
\end{equation}
where $P$ and $Q$ are the true and expressed opinion distributions, $M=\frac{1}{2}(P+Q)$, and 
\begin{equation}
D(P||Q)=\sum_i P(i) \log \frac{P(i)}{Q(i)}.
\end{equation}
The JSD ranges from $0$ (identity) to $1$ (minimum mutual information)


\section*{Results}
\addcontentsline{toc}{section}{Results}

\subsection*{Experiment 1: Social Influence and Intolerance}
\begin{figure*}
\centering
\includegraphics[width=0.49\textwidth]{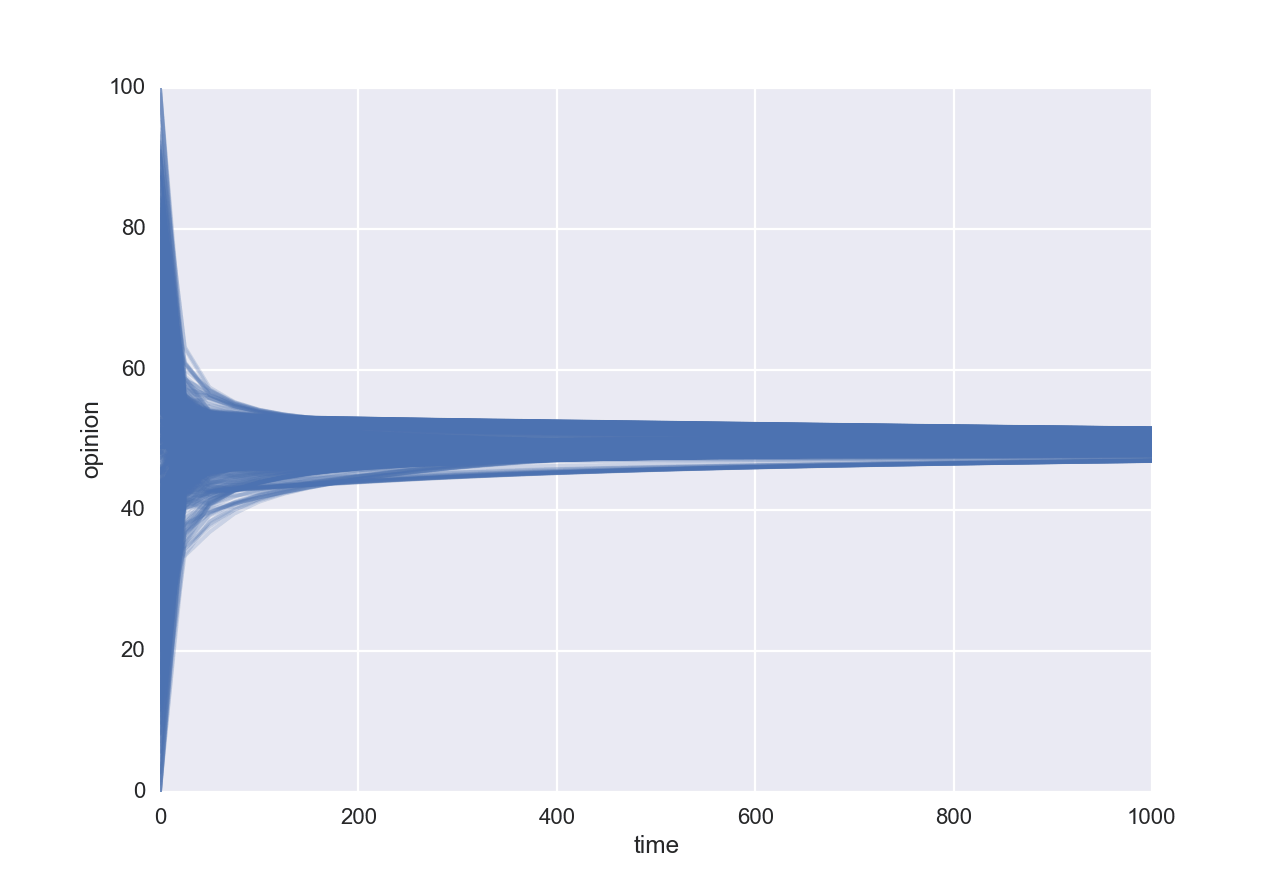}
\includegraphics[width=0.49\textwidth]{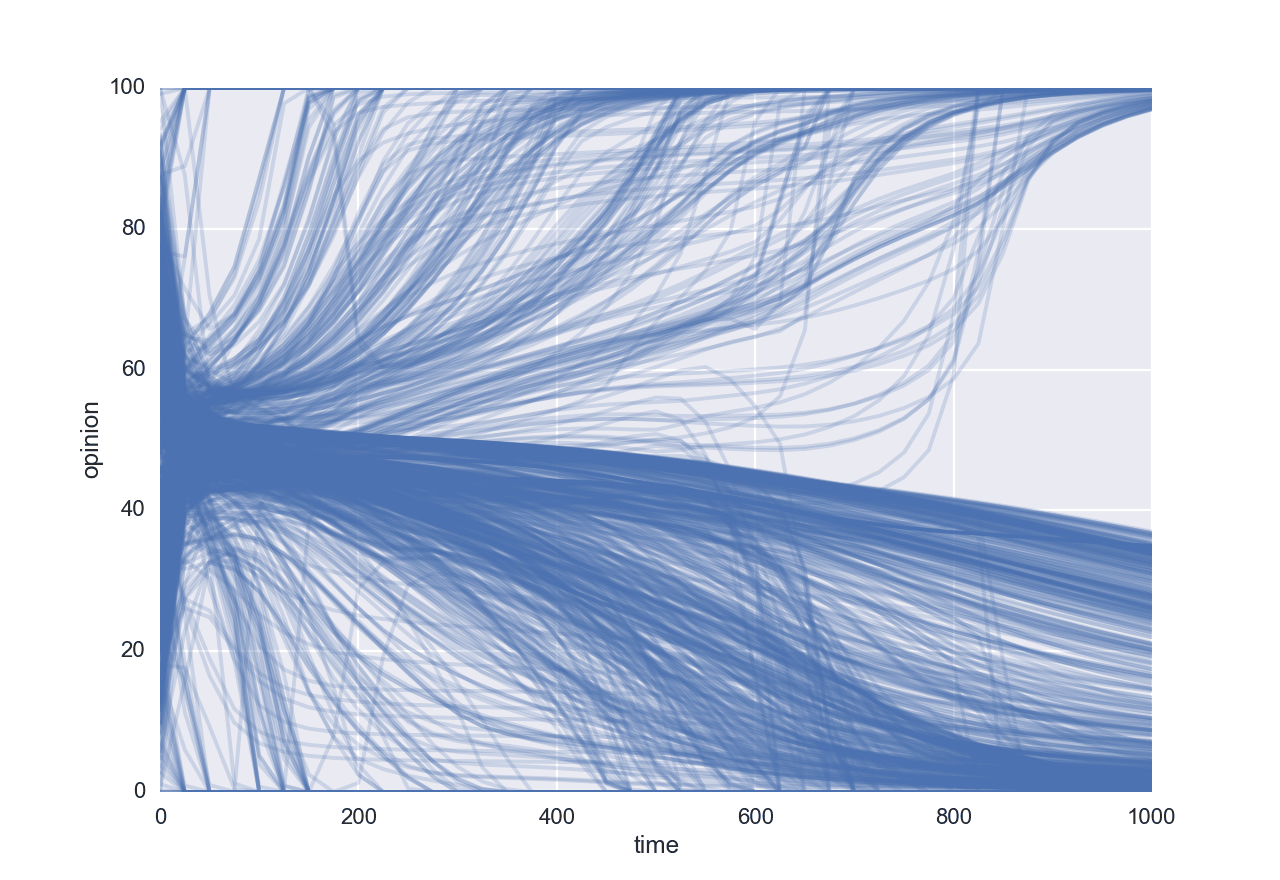}
\caption{A society populated by agents who initially hold normally distributed opinions converges to a single centrist opinion if agents' intolerance of dissimilar opinions is low, and diverges to two extremists opinion if agents' intolerance is high. This result reproduces findings in classical opinion dynamics and represents the base case of the simulation, in which social influence is the only active psychological force and all agents are identically intolerant. Additional runs show that societies with intolerance below $\mu_t=0.7$ always converge, societies with intolerance above $\mu_t=1.0$ always diverge, and societies in between can either converge or diverge, depending on initial conditions. Strong diversity doesn't emerge.}
\label{tolerance}
\end{figure*}

To begin, I reproduce a classical experiment in opinion dynamics, in which the final distribution of opinions is examined as a function of intolerance. In this model, intolerance is an agent-level parameter $t_i$ which is initially drawn from a normal distribution of mean $\mu_t$ and variance $\sigma_t$. For these preliminary experiments I assume no heterogeneity of intolerance, susceptibility, or conformity: $\sigma_t,\sigma_s,\sigma_c=0$.
\begin{center}
\textbf{Hypothesis 1}: low intolerance promotes societal opinion convergence, while high intolerance produces opinion polarization and weak diversity.
\end{center}
In a society with low intolerance, $\mu_t=0.7$, most agents assign positive weight to each others' opinions during dialogues, and are consequently pulled towards the mean opinion in that dialogue. Figure \ref{tolerance} (left) shows that an initial normal distribution of opinions, $\mu_O=50,\sigma_O=20$, rapidly converges to a single, centrist opinion: given enough time, diversity will completely disappear, and all agents will believe $O_i=50$. Conversely, in a society with high intolerance, $\mu_t=1.0$, many agents assign negative weight to each others' opinions and are pushed away from the dialogue mean. As agents adopt stronger opinions, they assign stronger negative weights, resulting in polarizing feedback. Figure \ref{tolerance} (right) shows this society rapidly diverges to two extremists opinions at either end of the opinion spectrum. As $t\rightarrow \infty$ only weak diversity remains: all agents either hold $O_i=0$ or $O_i=100$. These base-case results confirm the classical finding that, in the absence of other psychological forces, the degree of individuals' intolerance determines whether society homogenizes or polarizes.

\subsection*{Experiment 2: Conformity and Distinctiveness}
Next I introduce social context into the simulation by allowing agents to misrepresent their true opinions in dialogues. Though opinion falsification does not directly affect agents' true opinion update, it does affect the information available to those agents. If falsification is significant, agents will perceive an unrepresentative distribution of opinions (compared to each others' true beliefs) and change their beliefs accordingly.

\begin{center}
\textbf{Hypothesis 2}: a conformist society will homogenize under conditions that otherwise cause polarization, while a society driven by distinctiveness will polarize under conditions that otherwise favor consensus.
\end{center}

\begin{figure*}
\centering
\includegraphics[width=0.49\textwidth]{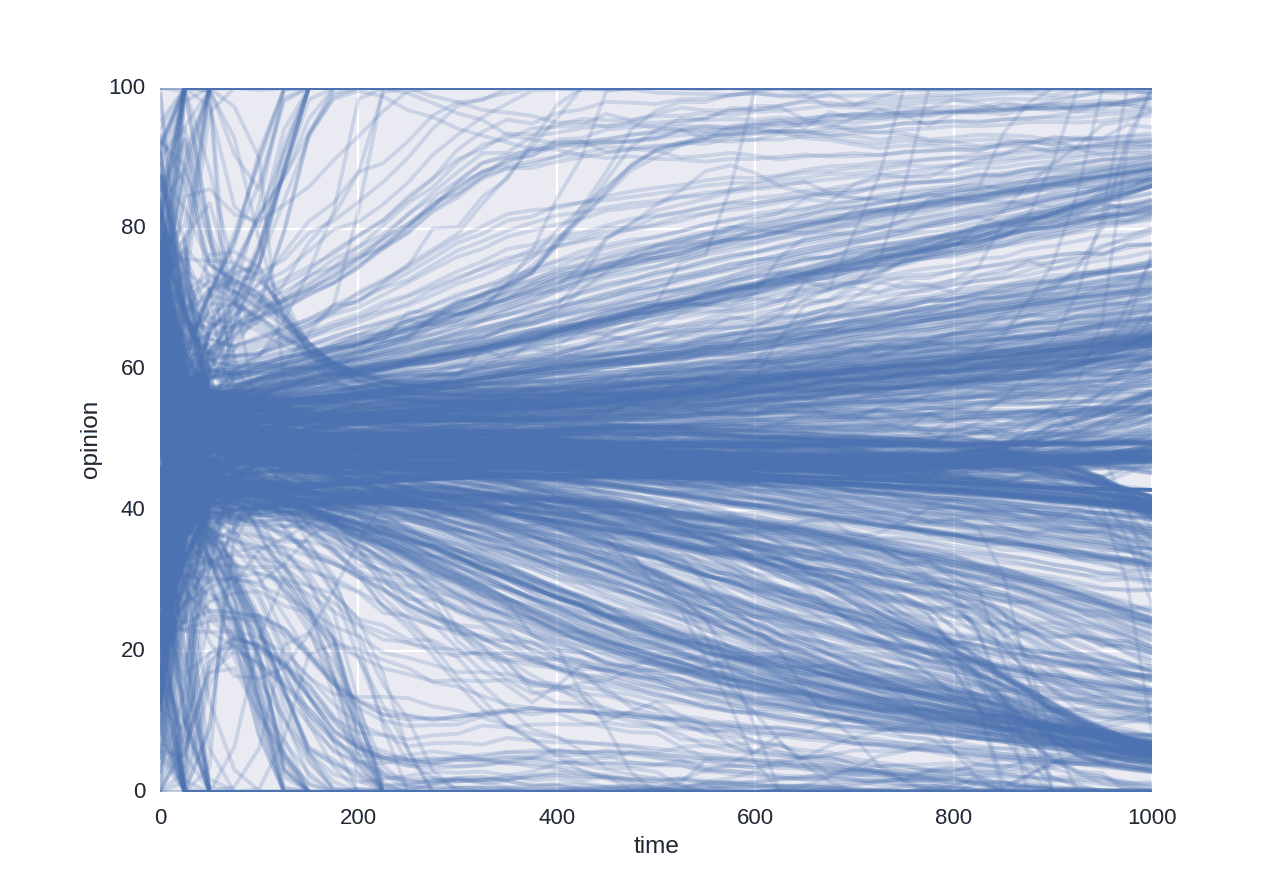}
\includegraphics[width=0.49\textwidth]{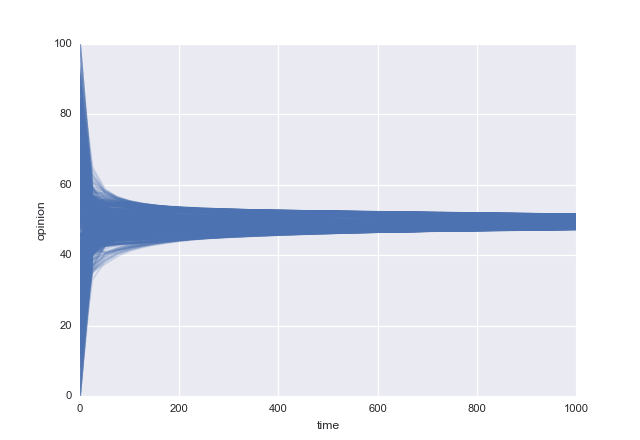}
\caption{When agents with low intolerance wish to appear distinct in social contexts, they express extremists beliefs in dialogues, which eventually polarizes society and leaves weak, bimodal diversity. When agents with high intolerance are motivated to socially conform, they express normative centrist views, causing society to converge to centrism.}
\label{conformity}
\end{figure*}

First, I simulate a society whose high intolerance would normally cause polarization, $\mu_t=1.0$, but introduce a moderate tendency towards social conformity, $\mu_c=0.5$. Conforming agents now express opinions that are close enough to the dialogue mean that almost nobody assigns these (falsified) opinions a negative weight. Agents adopt opinions closer to the norms expressed in the dialogue, and opinions converge, eventually resulting in societal consensus as shown in Figure \ref{conformity} (right). Second, I simulate the opposite conditions: a society whose low intolerance would normally cause convergence, $\mu_t=0.7$, but filled with agents who possess a strong desire to be distinct, $\mu_c=-1.5$. When agents converse, they express opinions that differ radically from centrist norms. As expressions become more extreme, social influence causes agents to adopt and retain extreme beliefs. Figure \ref{conformity} (left) shows that opinions initially converge, then diverge towards two homogeneous extremist parties. These results indicate that contextual opinion falsification can reverse the effects of intolerance, but cannot sustain strong diversity or pluralistic ignorance. 

\subsection*{Experiment 3: Commitment to Strong Beliefs}
I conclude the preliminary experiments by investigating whether, when agents' susceptibility to influence decreases with their belief extremity, different patterns of societal opinion change emerge.
\begin{center}
\textbf{Hypothesis 3}: when extremist agents undergo less opinion change than moderate or centrist agents, their persistent influence will prevent centrist homogenization and produce weak diversity.
\end{center}
Beginning with the simple case of a tolerant society with no opinion falsification, I test whether strong commitment, $\mu_s=10.0$, can reverse trends towards convergence. The mean opinion expressed in dialogues is still $\bar{O}\simeq50$, but because agents are tolerant and truthful, they assign positive weights to all opinions they hear. Extreme agents undergo little opinion change due to their commitment, but without a repelling force to push them away from social norms (i.e. intolerance or distinctiveness), they are still pulled slowly towards this centrist opinion. Although they remain steadfast in their views for longer periods of time than in Experiment 1, they eventually converge to a single centrist opinion like the rest of society (not shown). This result contradicts Hypothesis 3, showing that commitment by itself cannot reverse homophilous opinion convergence.

\begin{figure*}[p]
\centering
\includegraphics[width=1.0\textwidth]{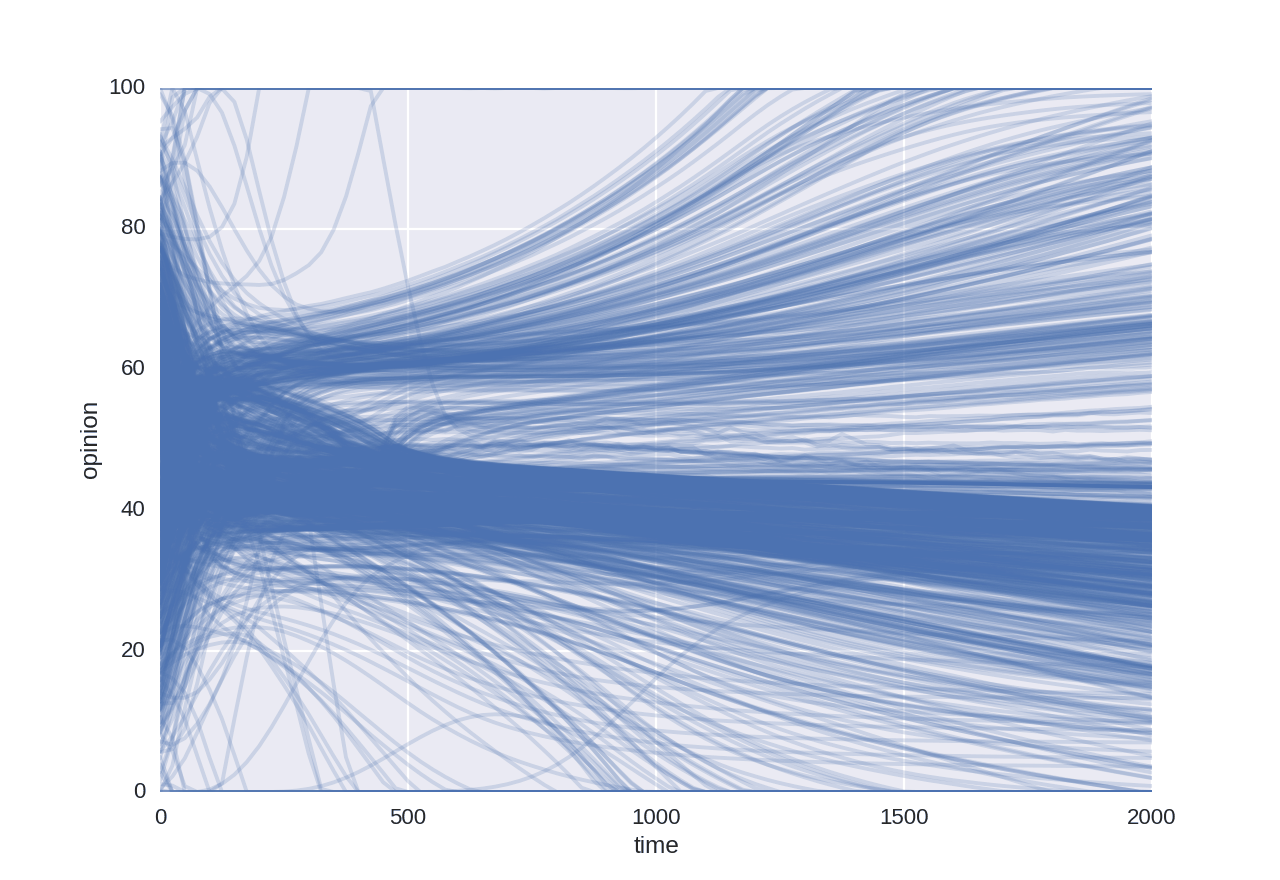}
\includegraphics[width=0.49\textwidth]{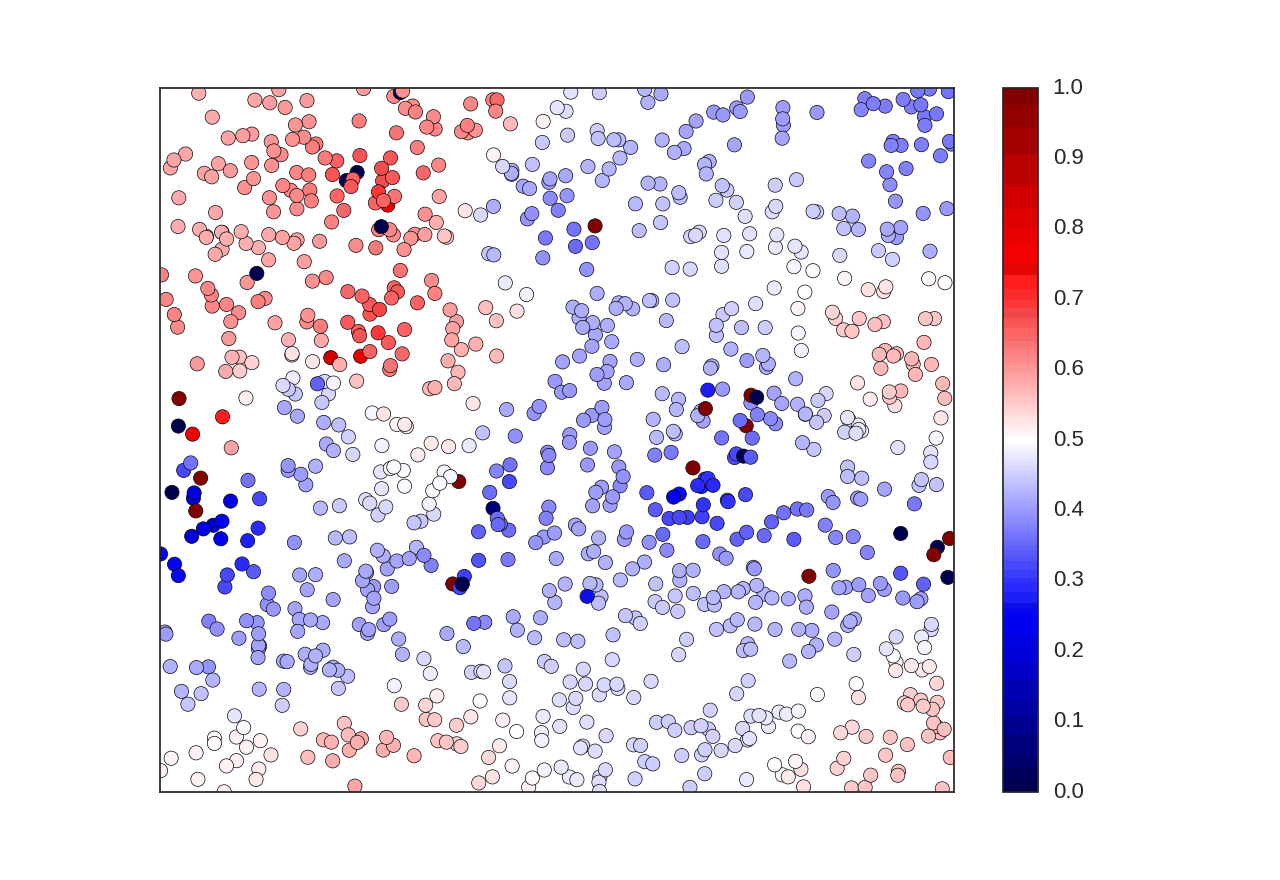}
\includegraphics[width=0.49\textwidth]{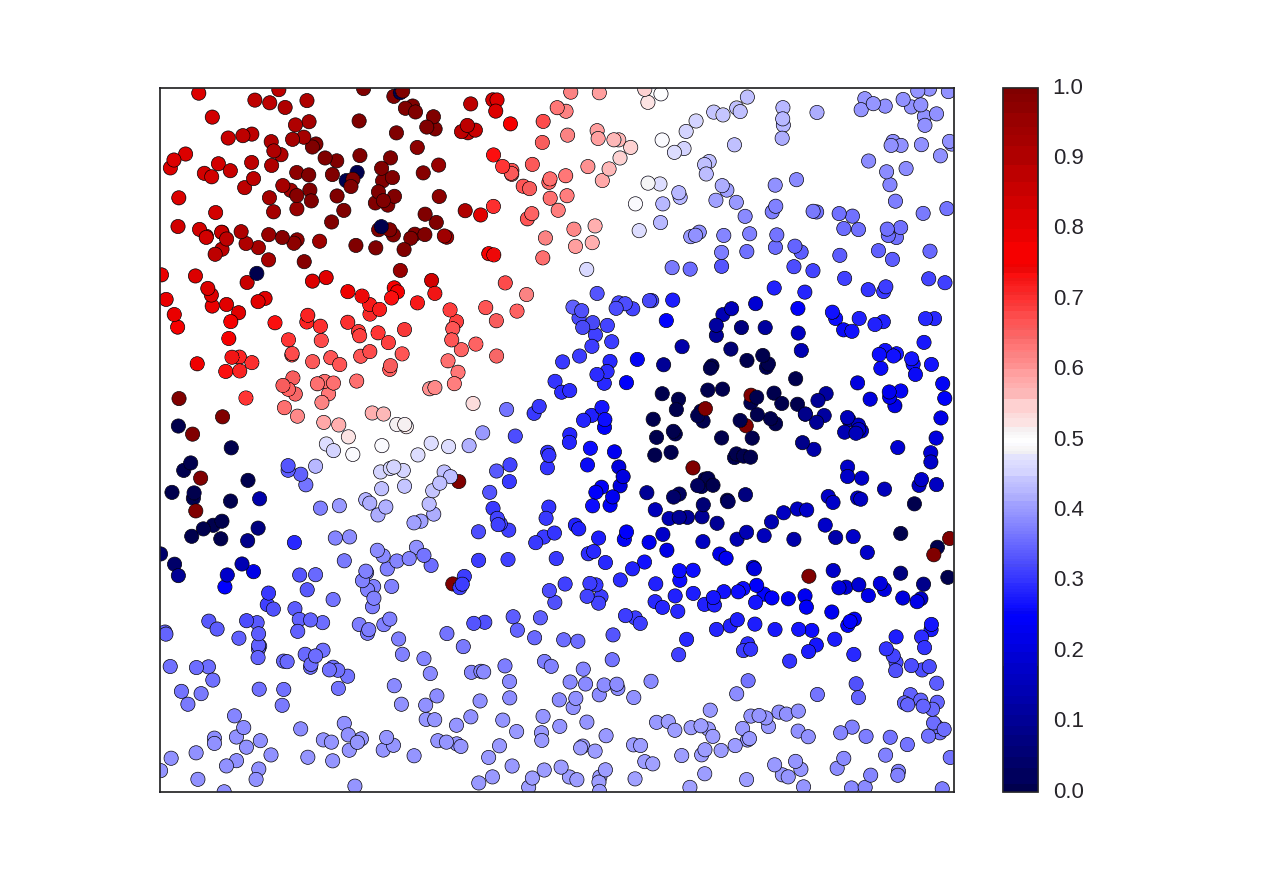}
\caption{In a society filled with individuals who are intolerant, conformist, and committed to extreme beliefs, a minority of agents will distance themselves from centrist norms and stubbornly express extremist views. The influence of these agents polarizes their neighbors, spreading extremism spatially outward until society bifurcates into weak diversity. Spatial maps at $t=200$ and $t=2000$ show that polarization originates from neighborhoods that contain extremist agents.}
\label{susceptibility}
\end{figure*}

However, personal susceptibility can affect a society that is intolerant and conformist, which normally homogenizes as in Experiment 2. Opinions initially converge due to the strong centrist norms perceived in conformist social dialogues, but extremists are slow to change. By $t=300$, most agents have adopted moderate or centrist opinions and expressions, but about $2\%$ of agents have, through intolerant repelling, adopted maximally extreme opinions, as shown in Figure \ref{susceptibility}. These extreme agents are now so committed to their beliefs that they barely soften their expressions to socially conform, $O_i=100 \rightarrow E_i=95$, and their strongly opinionated vocalizations polarize their social networks. Over time, this influence bifurcates society, as can be seen by the divergence of opinions past $t=500$. This experiment indicates that personal commitment fosters pockets of extremism whose long-term influence significantly alters societal opinion dynamics. 

\subsection*{Experiment 4: Strong Diversity}
Equipped with a basic understanding of model behavior and the independent effects of intolerance, susceptibility, and commitment, I now simulate societies in which all three psychosocial forces interact. In this experiment, all agent parameters are drawn from normal distributions with nonzero means and variances, creating an artificial society with greater heterogeneity and psychological realism than previous opinion dynamics models.
\begin{center}
\textbf{Hypothesis 4}: when agents are simultaneously motivated by social influence, personal susceptibility, and social context of varying degrees, society will (a) maintain a strong diversity of opinions and (b) exhibit and pluralistic ignorance.
\end{center}

\begin{figure*}[p]
\centering
\includegraphics[width=1.0\textwidth]{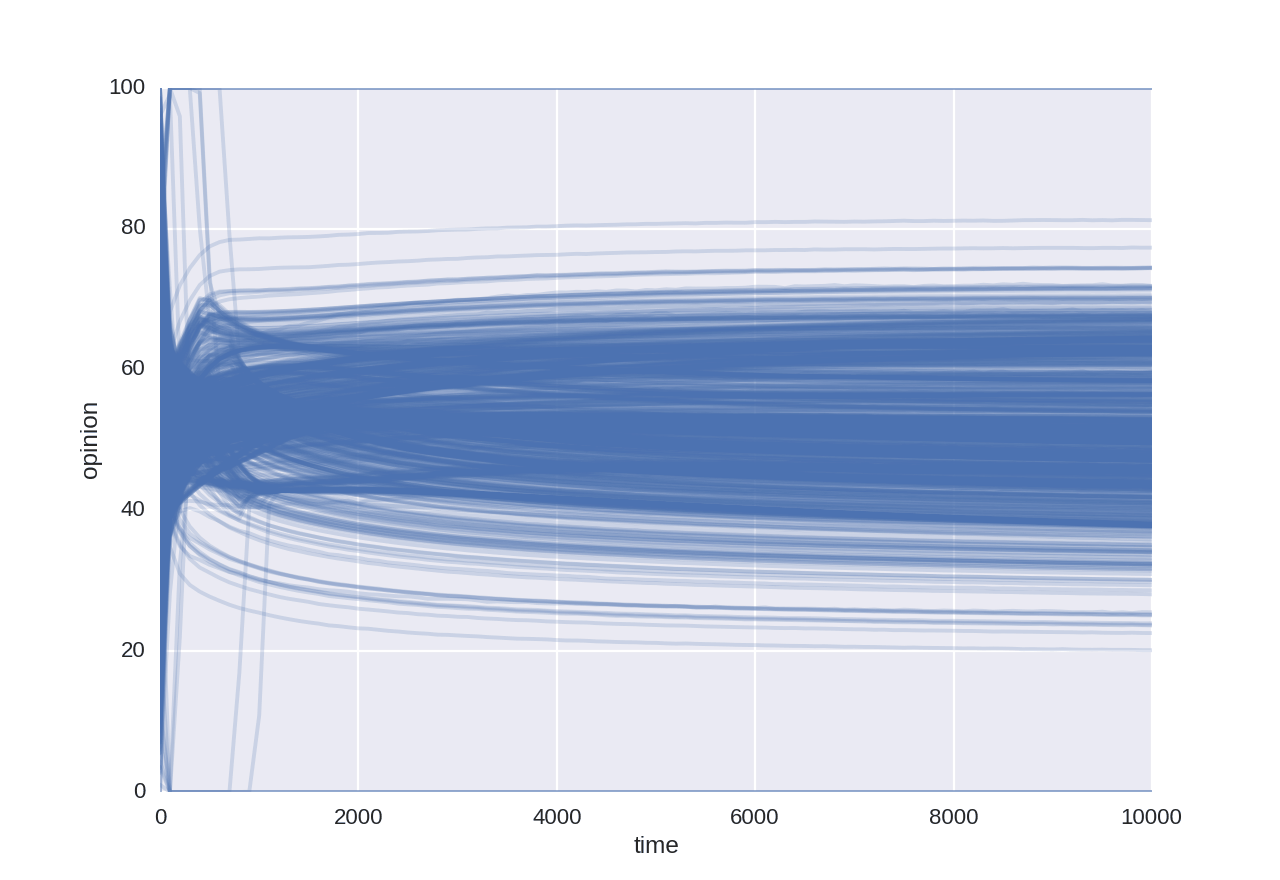}
\includegraphics[width=0.49\textwidth]{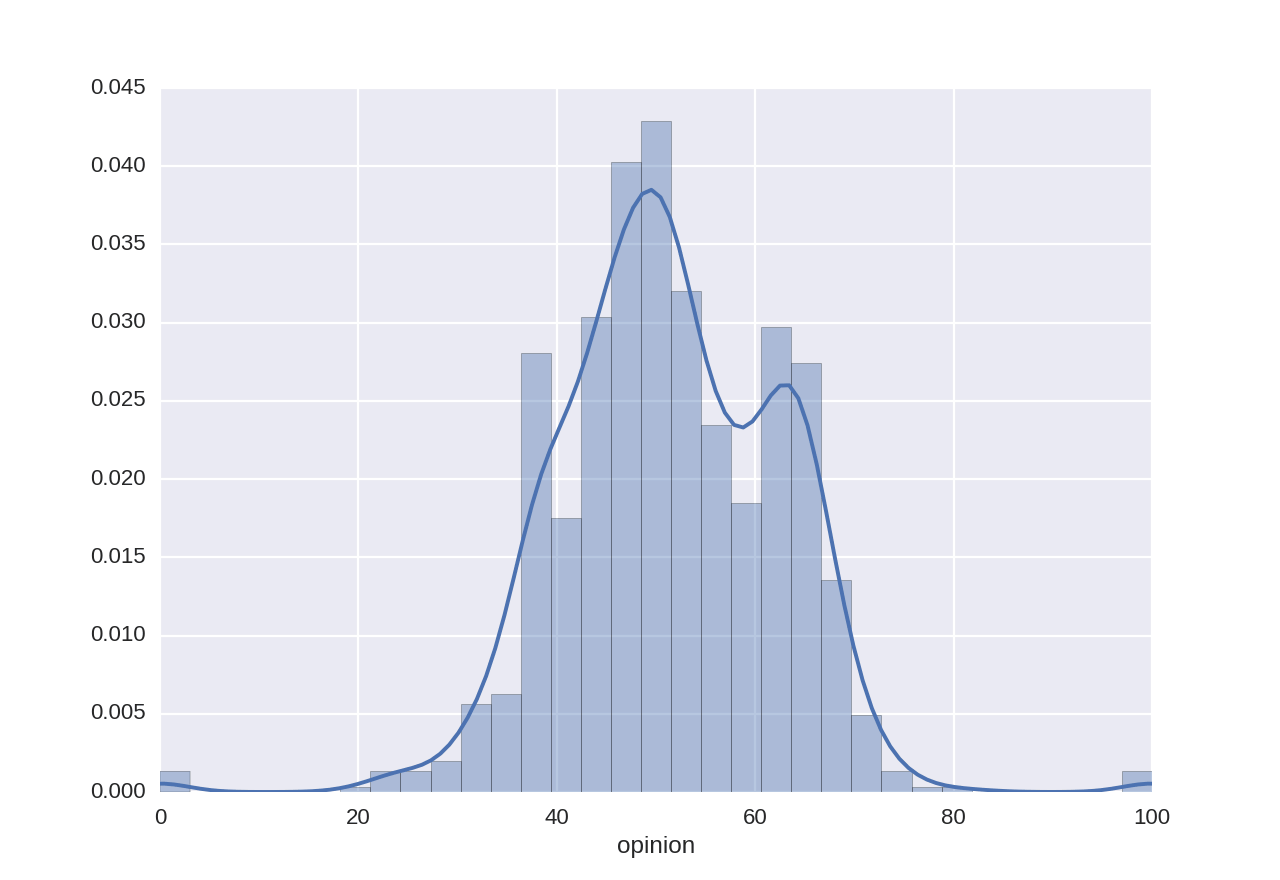}
\includegraphics[width=0.49\textwidth]{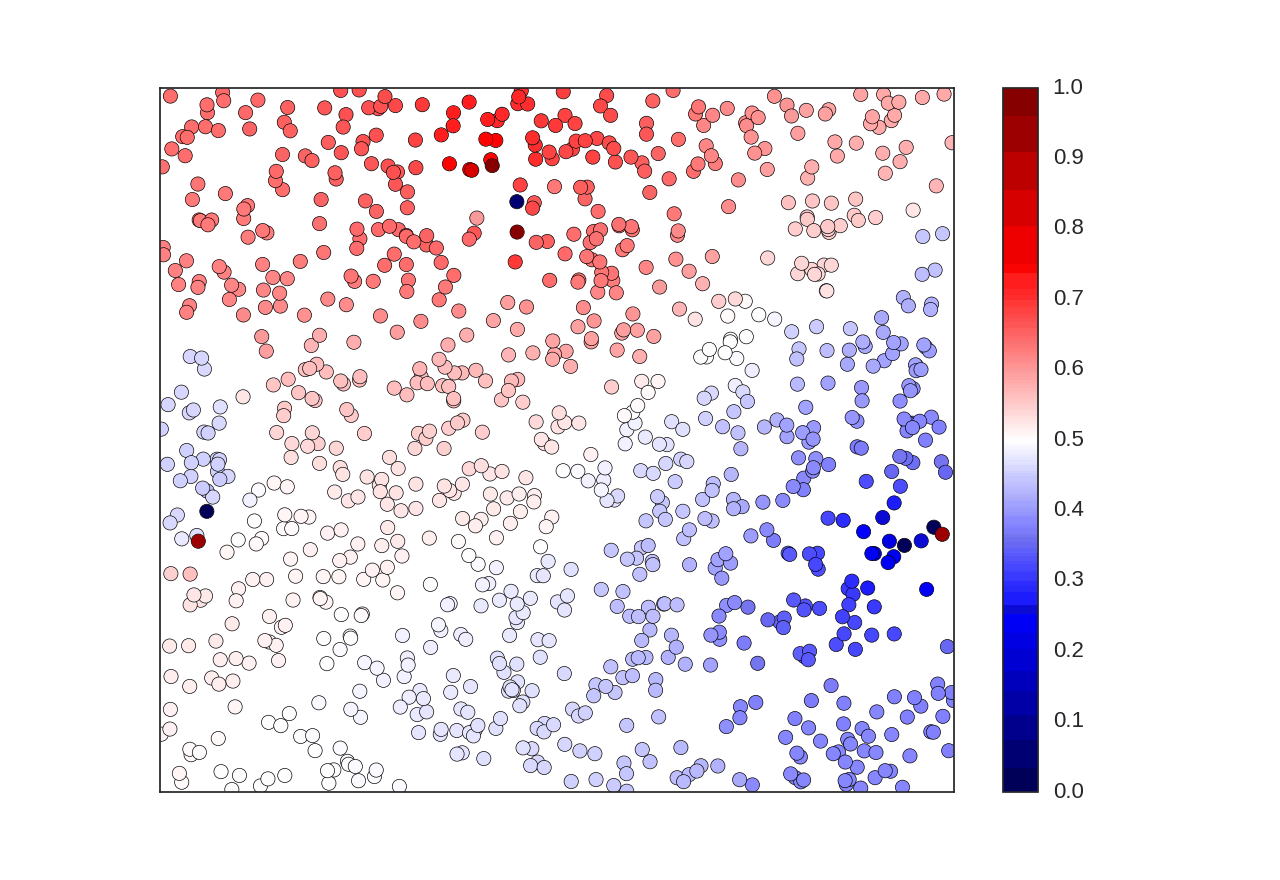}
\caption{When agents are motivated by social influence, personal susceptibility, and social context of heterogeneous strength, novel opinion distributions emerge at the societal scale. The opinion trajectory shows that society settles into a stable opinion configuration, while the opinion histogram confirms that the final distribution of opinions is strongly diverse. As confirmed by the spatial map, most agents have adopted a centrist opinion, but a small minority of extremists counterbalance homogenizing norms, preventing total convergence but exerting too little influence to bifurcate society}.
\label{diversity}
\end{figure*}

First I examine a society which is on average tolerant and conformist ($\mu_t=0.7, \mu_c=0.4$), but with enough diversity to create some intolerant and distinction-seeking agents ($\sigma_t=0.3, \sigma_c=0.1$). These forces tend to pull society towards convergence. However, an opposing commitment to strong beliefs ($\mu_s=5.0,\sigma_s=0.4$) helps initially-extreme agents resist normalization and locally exert polarizing influence. The opinion trajectory plot in Figure \ref{diversity} shows that society initially converges, but scattered extremists retain their strong views. Unlike in Experiment 3, where strong commitment kept extremists from softening their expressions, extremists in this society moderate their expressions and few extreme opinions are publicly voiced. This, combined with low extremist density, prevents radicals from attracting many followers: by $t=1,000$, less than $5\%$ of the population holds or expresses extreme views. However, these extremists exert enough influence that they keep centrist norms from completely homogenizing society. By $t=10,000$, the distribution of opinions and expressions have settled into a diverse centrist group, a moderate fringe, and scattered extremists. This strongly diverse distribution of opinions persists past $t=100,000$ despite small opinion fluctuations. A spatial map of opinions shows that the diversity within the centrist party arises from the minor influence exerted by extremists, which keeps the surrounding neighborhoods to the ideological left and right of $\bar{O}\simeq50$. This result is, to my knowledge, the first evidence of indefinitely-sustained strong diversity in a continuous-opinion model.

\subsection*{Experiment 5: Opinion Subcultures}
\begin{figure*}[p]
\centering
\includegraphics[width=1.0\textwidth]{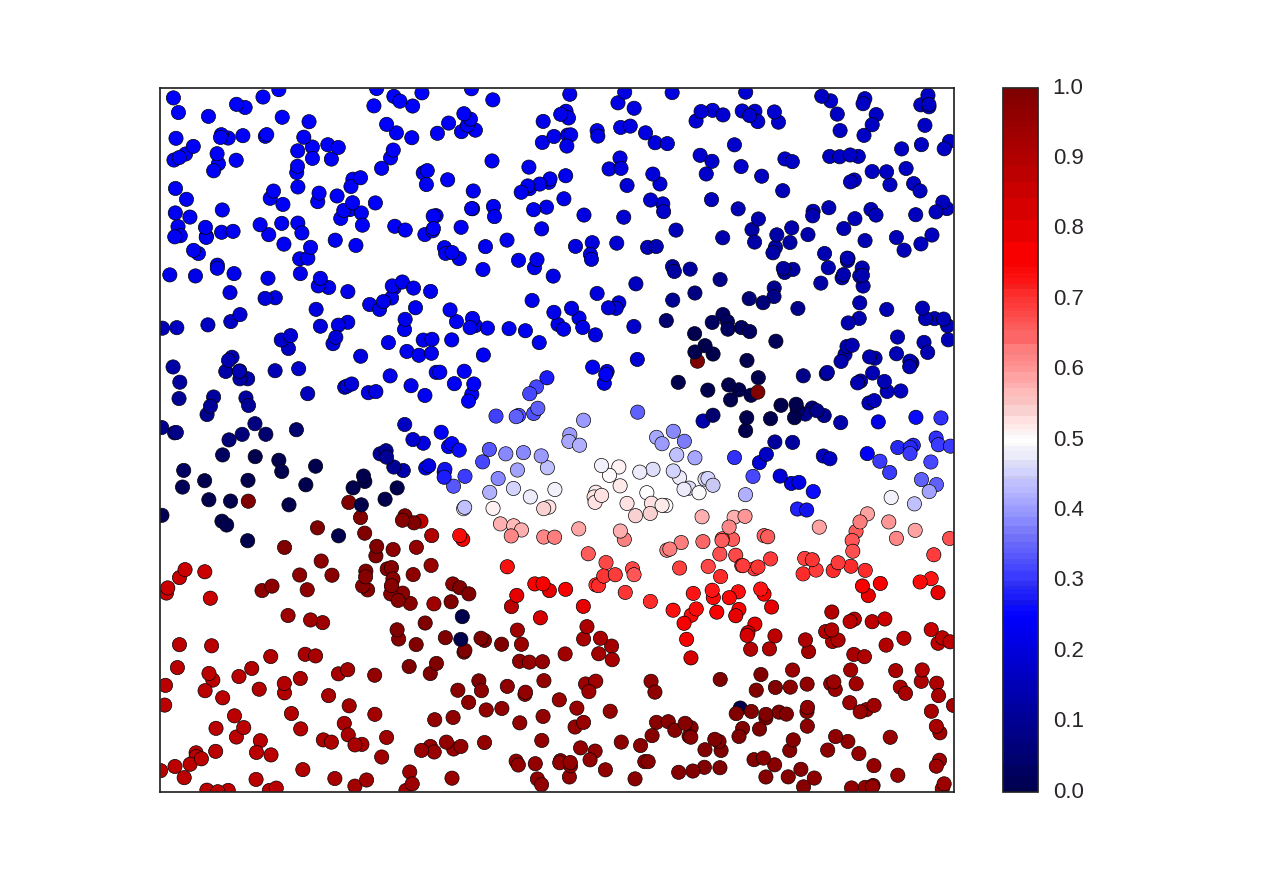}
\includegraphics[width=0.49\textwidth]{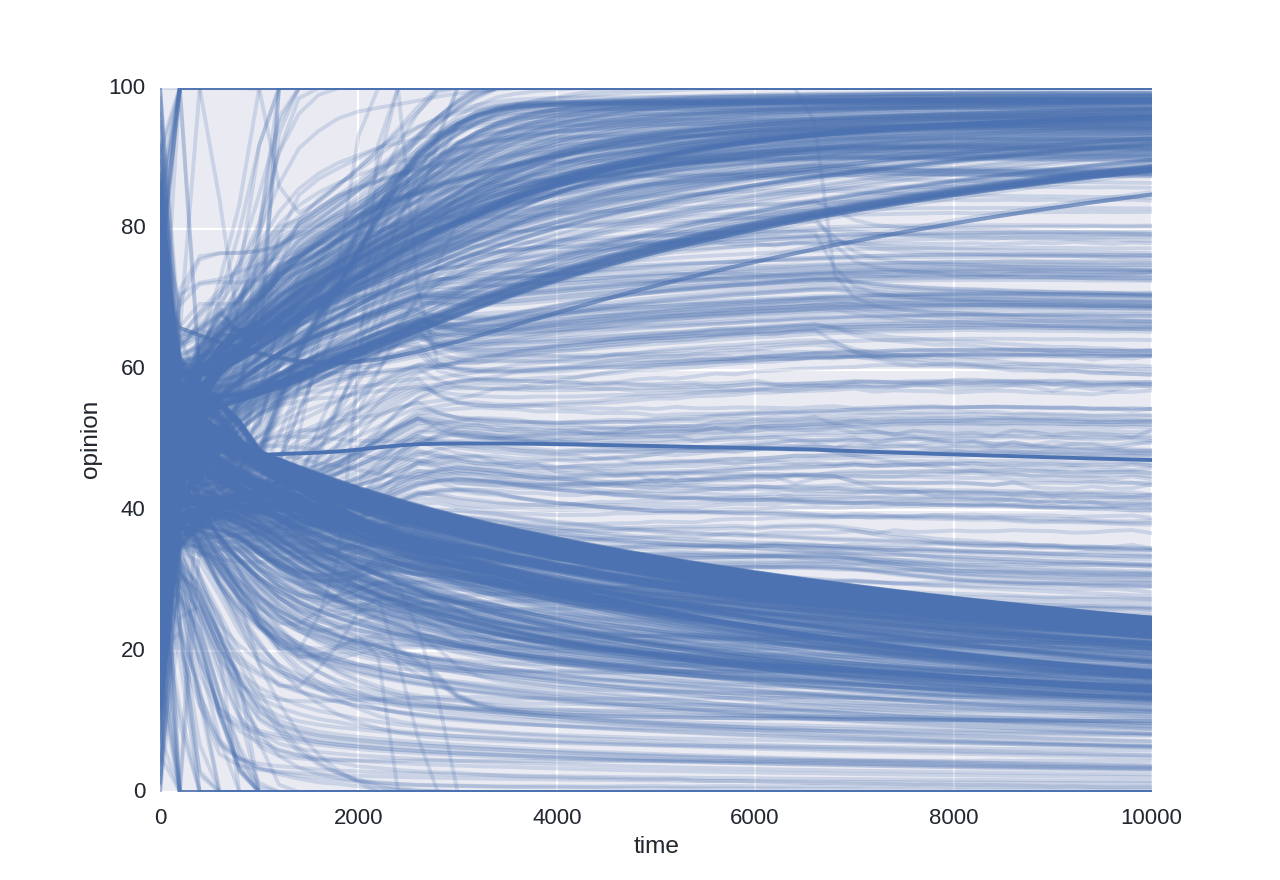}
\includegraphics[width=0.49\textwidth]{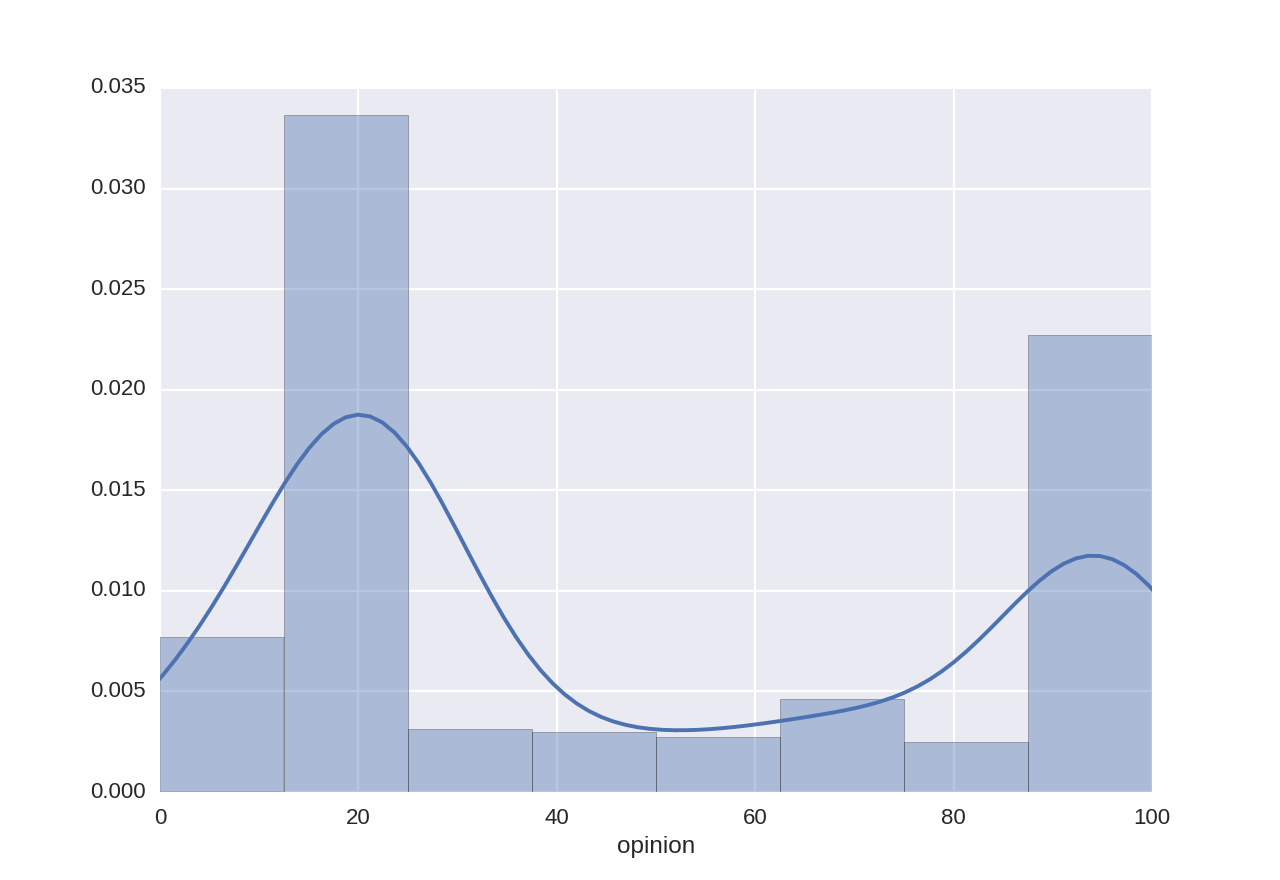}
\caption{Extremists counteract initial trends towards convergence and form neighborhoods of strongly (but uniquely) opinionated agents. These groups compete on the border: when one group exerts greater influence, they persuade moderate agents to become extremists; and when both groups exert equal influence, a buffer zone of centrist forms between them. These strongly diverse subcultures persist through time without artificial geographic or social barriers to prevent communication.}
\label{subcultures}
\end{figure*}

Next, I increase agents' average intolerance and conformity ($\mu_t=1.0$, $\mu_c=0.5$, $\mu_s=5.0$), then tweak their psychological diversity ($\sigma_t=0.3$, $\sigma_s=0.3$, $\sigma_c=0.3$). The results are shown in Figure \ref{subcultures}. After an initial period of convergence, several extremists neighborhoods develop, affecting partial polarization. Society quickly self-organizes into distinct, geographically-clustered opinion subcultures, as can be seen in the spatial map. These subcultures are stable and coherent, but continue to influence each other through persuadable agents on their mutual border. Eventually, society settles into two extremist groups and a centrist group. The spatial orientation of the extremist parties is such that the centrist party receives approximately equal influence from both sides of the opinion spectrum, and acts as a relatively stable buffer between the two extremes.

\subsection*{Experiment 6: Pluralistic Ignorance}
\begin{figure*}
\centering
\includegraphics[width=0.49\textwidth]{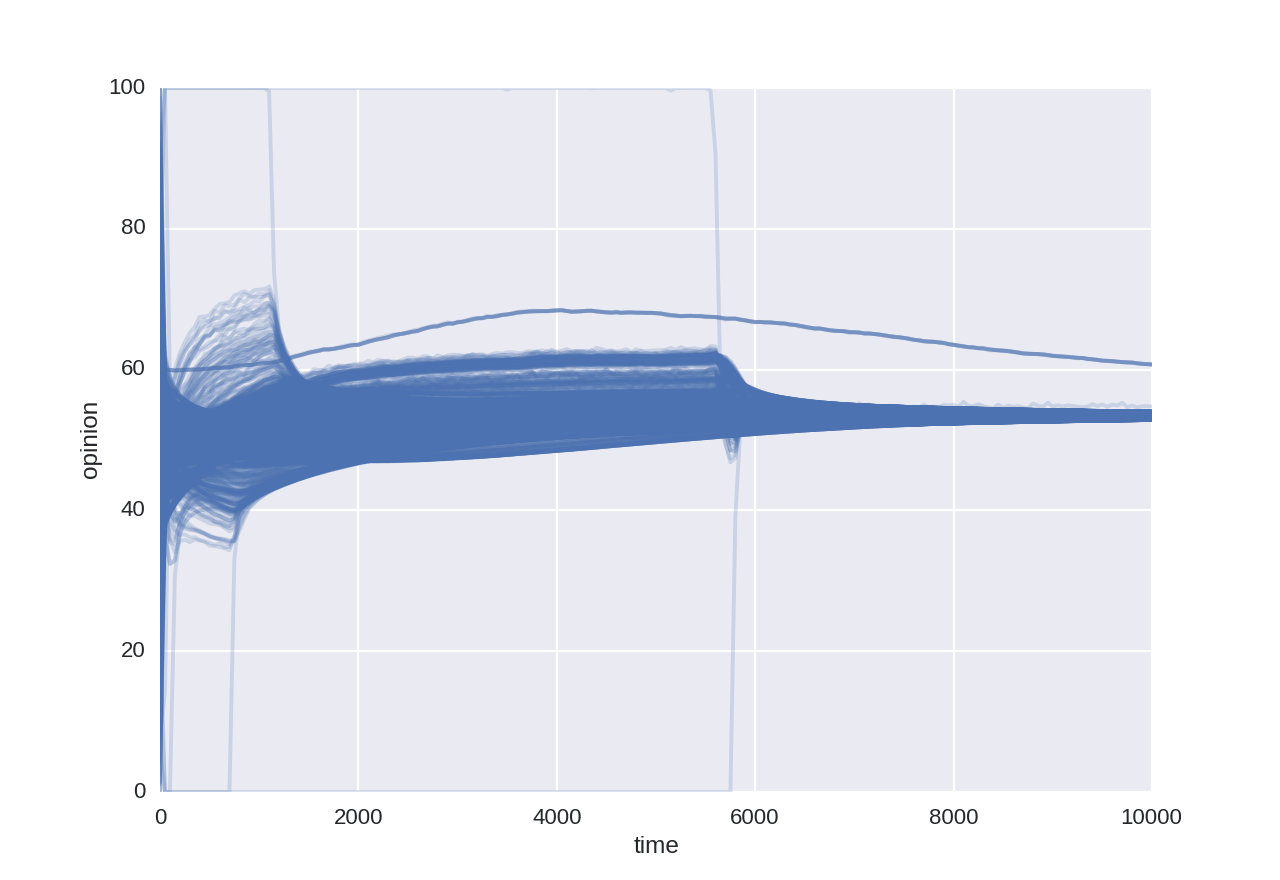}
\includegraphics[width=0.49\textwidth]{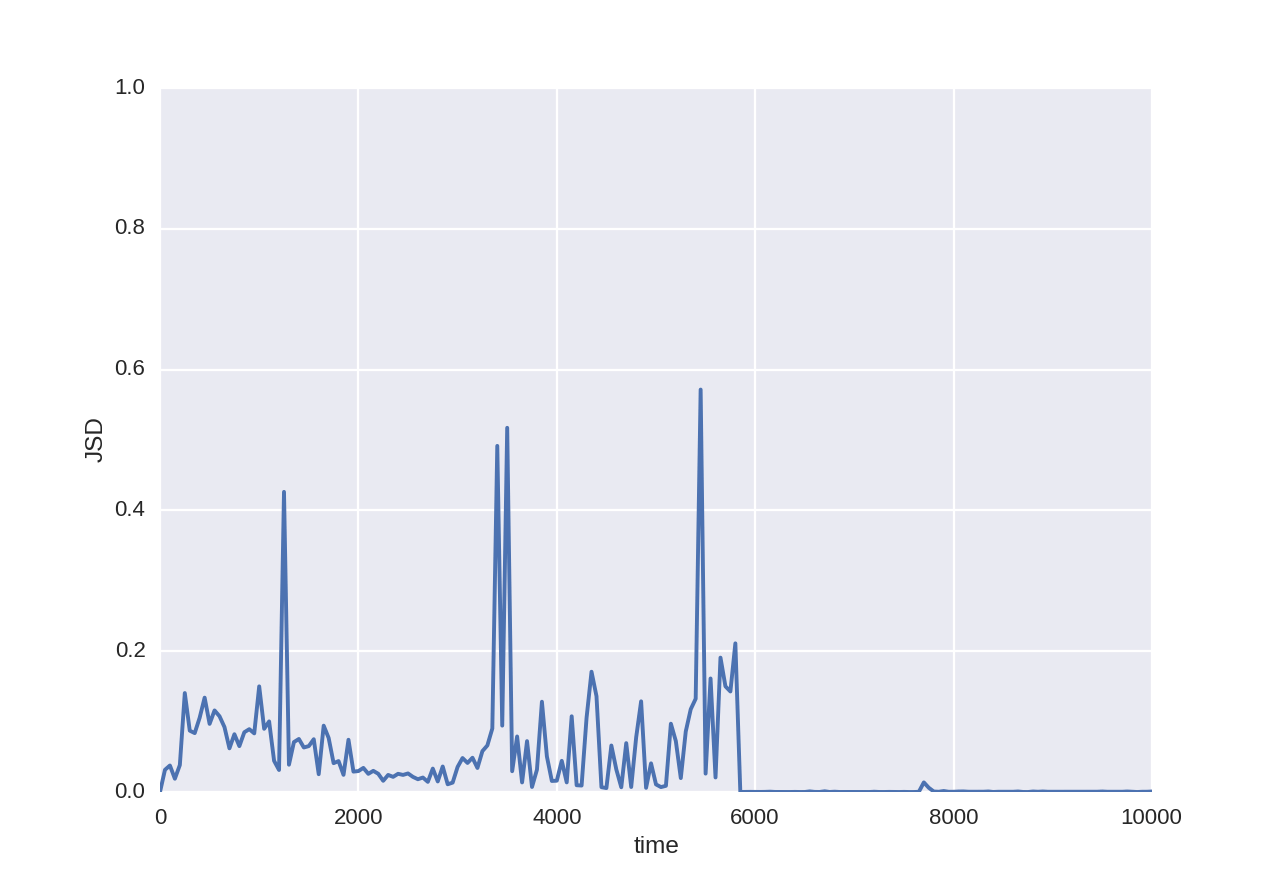}
\caption{The conversion of influential extremists may shift the course of societal opinion change from polarization to convergence. These events occur rapidly and are frequently preceded by spikes in pluralistic ignorance, such as at $t\simeq500,1000,5800$. This suggests that when sustained levels of opinion falsification are finally revealed, tipping-point phenomenon may occur, leading to nonlinear opinion dynamics.}
\label{pluralisticignorance}
\end{figure*}

Pluralistic ignorance and unpredictable dynamics are also possible under various conditions, such as when agents have intermediate intolerance ($\mu_t=0.8$, $\sigma_t=0.3$), low commitment ($\mu_s=0.1$, $\sigma_s=0.1$), and highly variable conformity ($\mu_c=0.3$, $\sigma_c=0.5$). Opinions converge early on, and society is sufficiently tolerant and uncommitted that only a few agents retain extreme opinions. Through some combination of the extremists' social influence, conformity of their neighbors, and distinction of agents from centrist norms, opinions throughout society begin drifting towards the extreme. However, unlike in Experiment 4, the extremists abruptly convert to centrism, causing a dramatic turn towards convergence, Figure \ref{pluralisticignorance}. Before these conversions, centrist or extremist agents express moderate opinions in dialogues, and pluralistic ignorance spikes. The perceived moderate norm pulls centrists towards extremism, causing the slow drift before $t=500$ in the opinion trajectories, but also pulls extremists towards centrism, causing the occasional conversion of an extremists agent. If the former trend dominates, society bifurcates; if the latter dominates, society homogenizes. Although this experiment shows that strong diversity does not always persist in the model, it suggests that pluralistic ignorance precedes dramatic and sometimes nonlinear changes in societal opinion dynamics.

\subsection*{Experiment 7: Social Reach}
Finally, I investigate whether strong diversity, opinion subcultures, and pluralistic ignorance are robust to changes in the size of agents' social networks:

\begin{center}
\textbf{Hypothesis 5}: small social networks will promote geographically-distinct opinion subcultures while large social networks will dissolve subgroups; strong diversity and pluralistic ignorance will not be affected.
\end{center}

\begin{figure*}[p]
\centering
\includegraphics[width=1.0\textwidth]{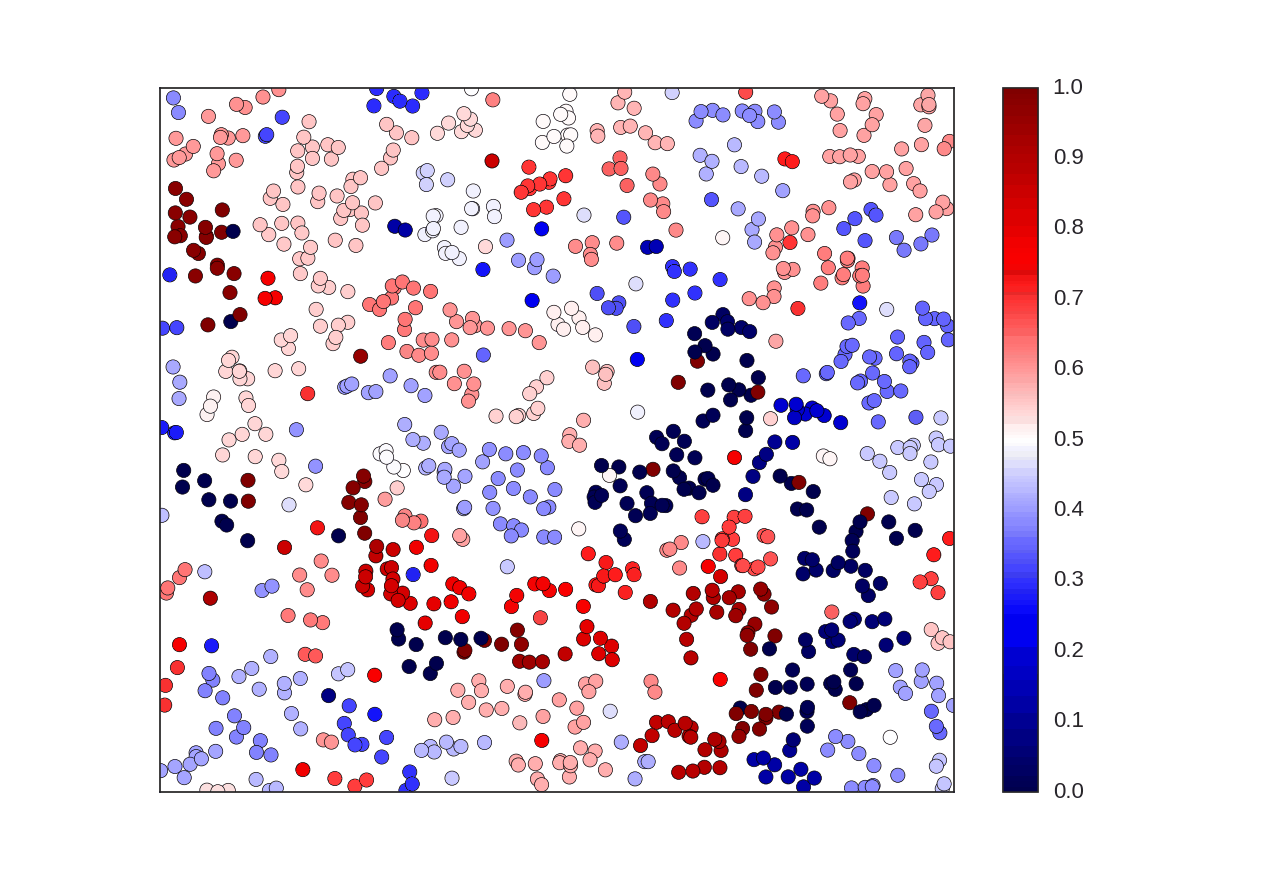}
\includegraphics[width=0.49\textwidth]{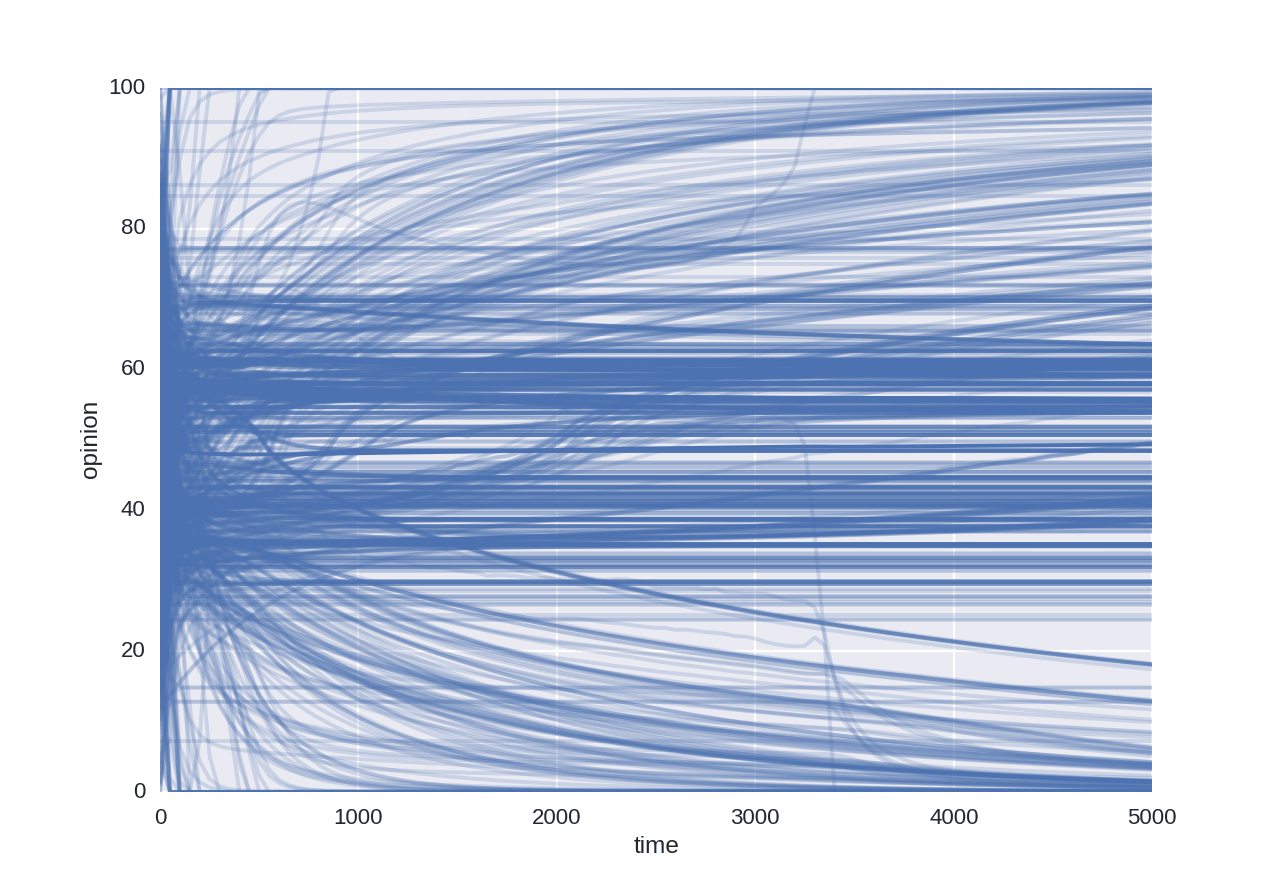}
\includegraphics[width=0.49\textwidth]{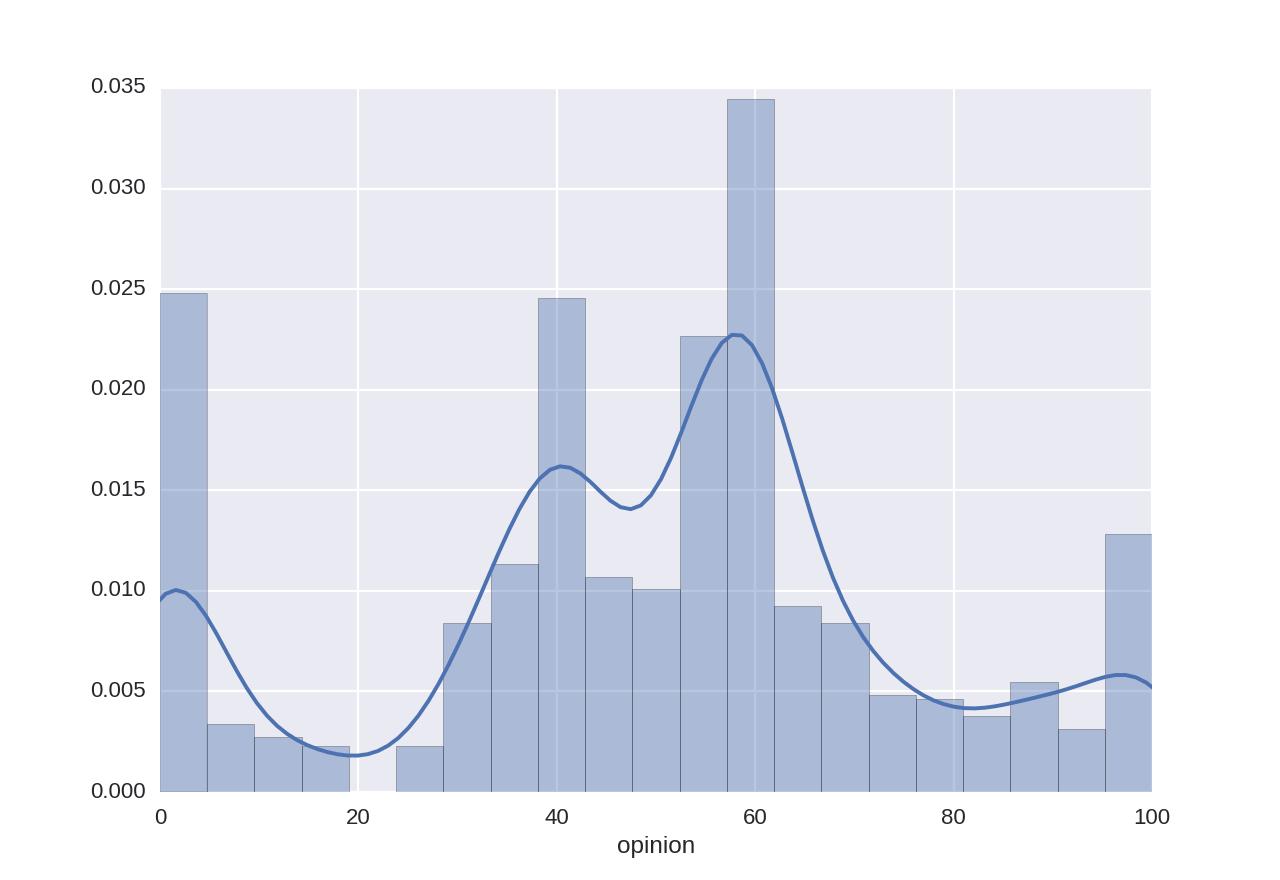}
\caption{Shrinking agents' social networks encourages the formation of geographically-organized opinion subcultures. Reduced interaction between these groups prevents both centrist and extremist takeover, but continuing dialogues with intermediary agents keeps these groups, and society as a whole, strongly diversity. Histograms and maps show opinions at $t=10,000$.}
\label{halfreach}
\end{figure*}

I reproduce Experiment 5 with smaller social networks, obtained by reducing agents' social reach (from $\mu_r=22$, $ \sigma_r=4$ to $\mu_r=11$, $ \sigma_r=2$). Opinions rapidly become clustered in geographically-constrained networks, producing discrete subcultures, Figure \ref{halfreach}. These subcultures continue to receive influence from surrounding networks, often through persuadable agents on the border who continually oppose consensus and promote strong diversity within each subculture. The partial isolation of subcultures prevents both homogenization and polarization globally, which is reflected in a wide, multimodal opinion distribution. Both opinion subcultures and strong diversity remain stable past $t=5,000$. Similar results were obtained by reducing social reach in Experiments 4 and 6.

\begin{figure*}[p]
\centering
\includegraphics[width=1.0\textwidth]{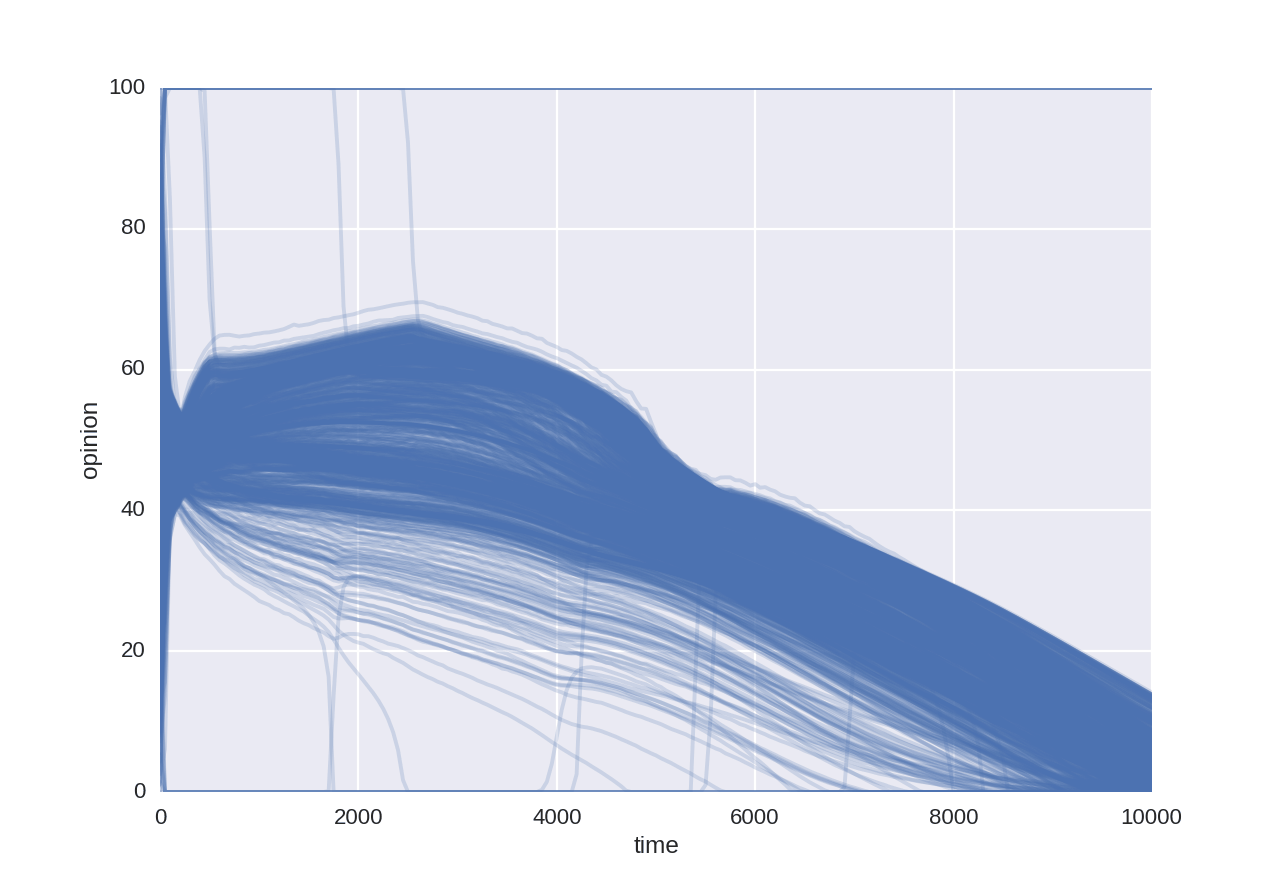}
\includegraphics[width=0.49\textwidth]{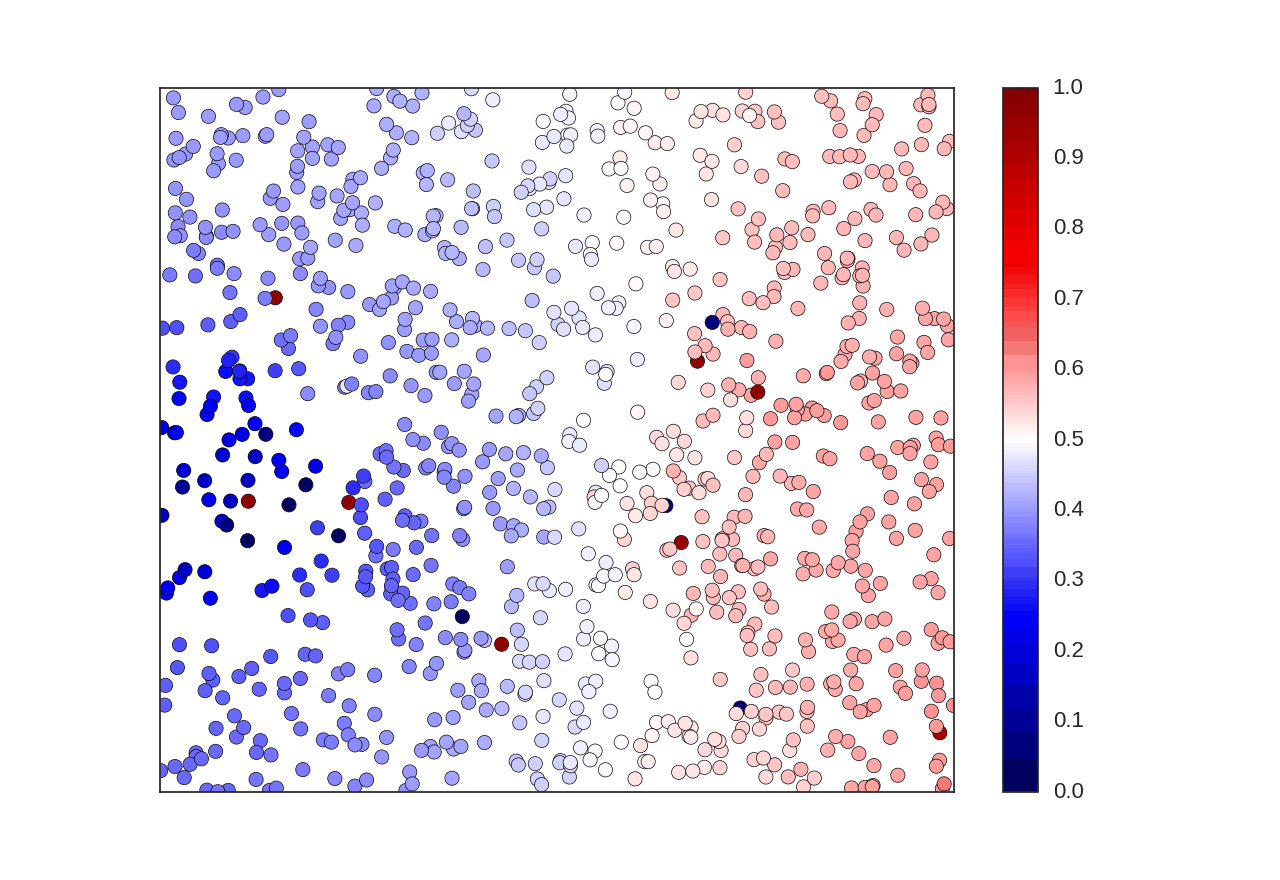}
\includegraphics[width=0.49\textwidth]{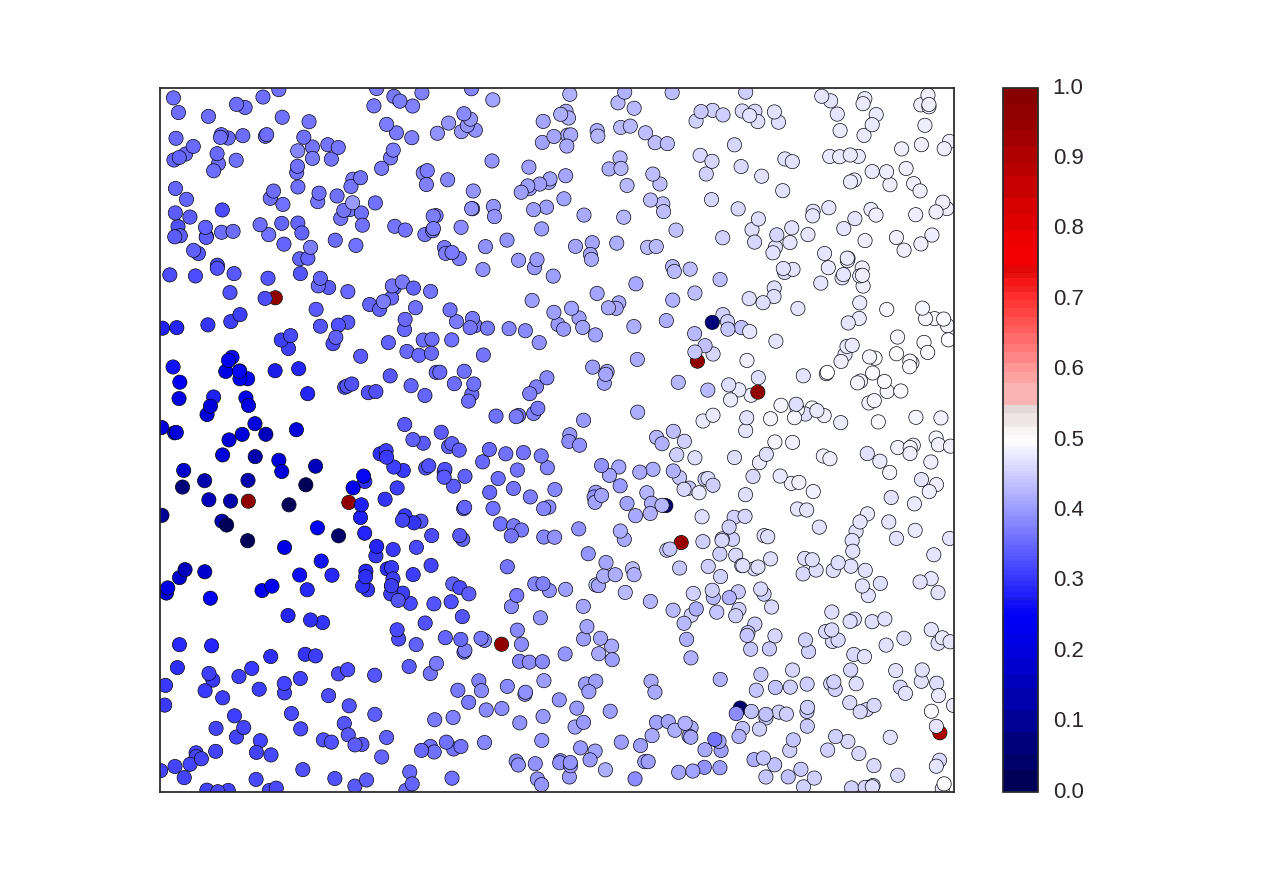}
\caption{Expanding agents' social networks dilutes the influence of extremist agents over larger network, while globalized influence prevents the formation of opinion subcultures and eventually destroys strong diversity. The spatial maps at $t=4,000$ and $t=5,000$ show that the buffer zone that previously preserved diversity no longer prevents the takeover of the dominant extremist group.}
\label{doublereach}
\end{figure*}

Conversely, increasing agents' social reach promotes homogenization. Initially, most agent converge to centrism, while the large size and strong centrist norms in dialogues encourage the remaining extremists to express moderate views. Two outcomes are possible: either all extremists convert and society converges to centrism; or, as shown in Figure \ref{doublereach}, an imbalance of extremists remains, and society drifts towards the most vocal group. Unlike in Experiment 5, where a buffer zone between extremist groups prevented takeover, large networks decrease the likelihood that centrists participate in dialogues that are well-balanced between the two extremes. This increases the probability that the dominant group will exert the strongest influence in all geographic regions, and that centrists and moderates will turn towards that extreme. In either case, strong diversity vanishes, in contradiction with Hypothesis 5. 

\section*{Validation: American Political Opinions}
\addcontentsline{toc}{section}{Validation: American Political Opinions}

Although the ISC model is grounded in social psychology and reproduces features of real-world opinion dynamics like strong diversity, opinion subcultures, and pluralistic ignorance, I have not shown that it quantitatively captures real-world data. In this section, I validate the model by reproducing empirical data on the distributions and dynamics of political opinions in American society. 

\begin{figure*}[p]
\centering
\includegraphics[width=0.49\textwidth]{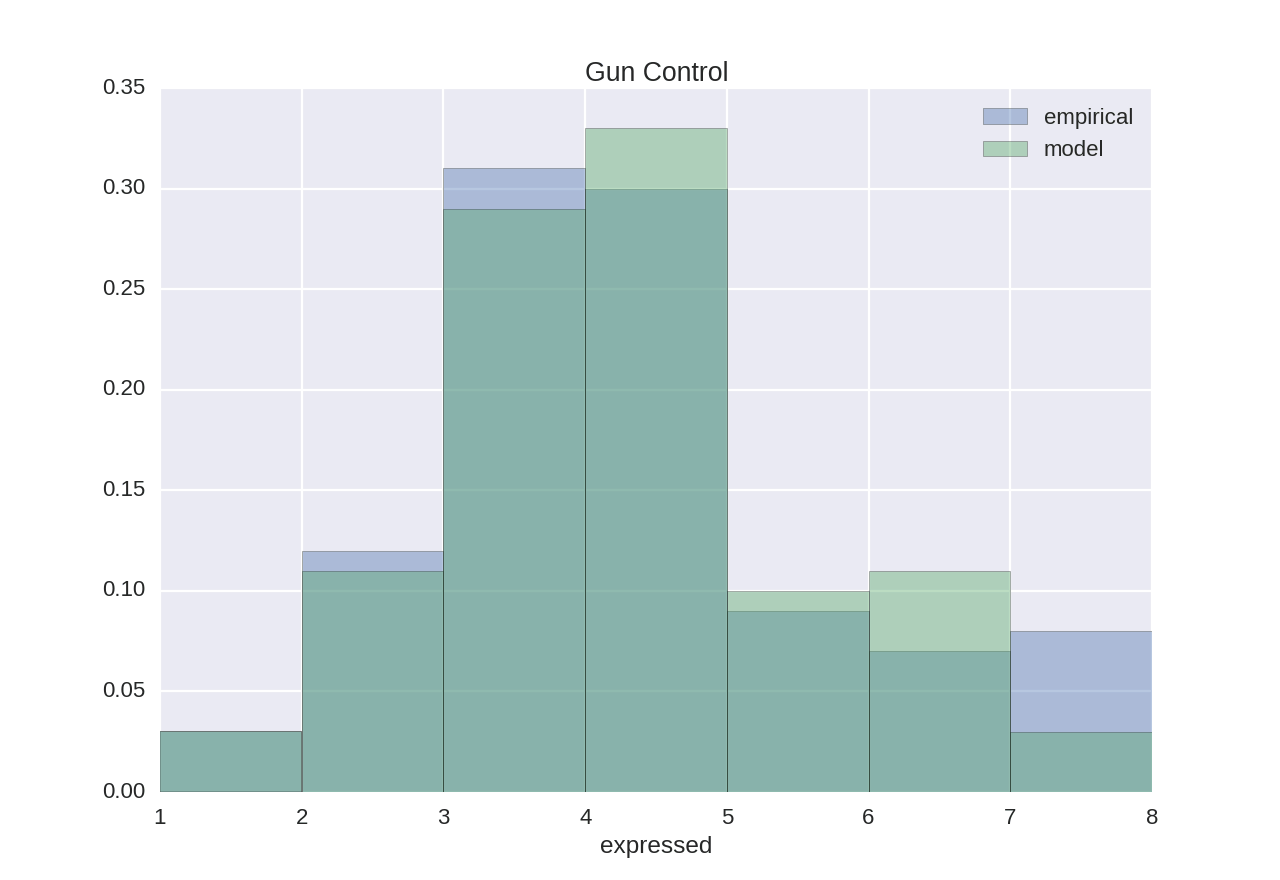}
\includegraphics[width=0.49\textwidth]{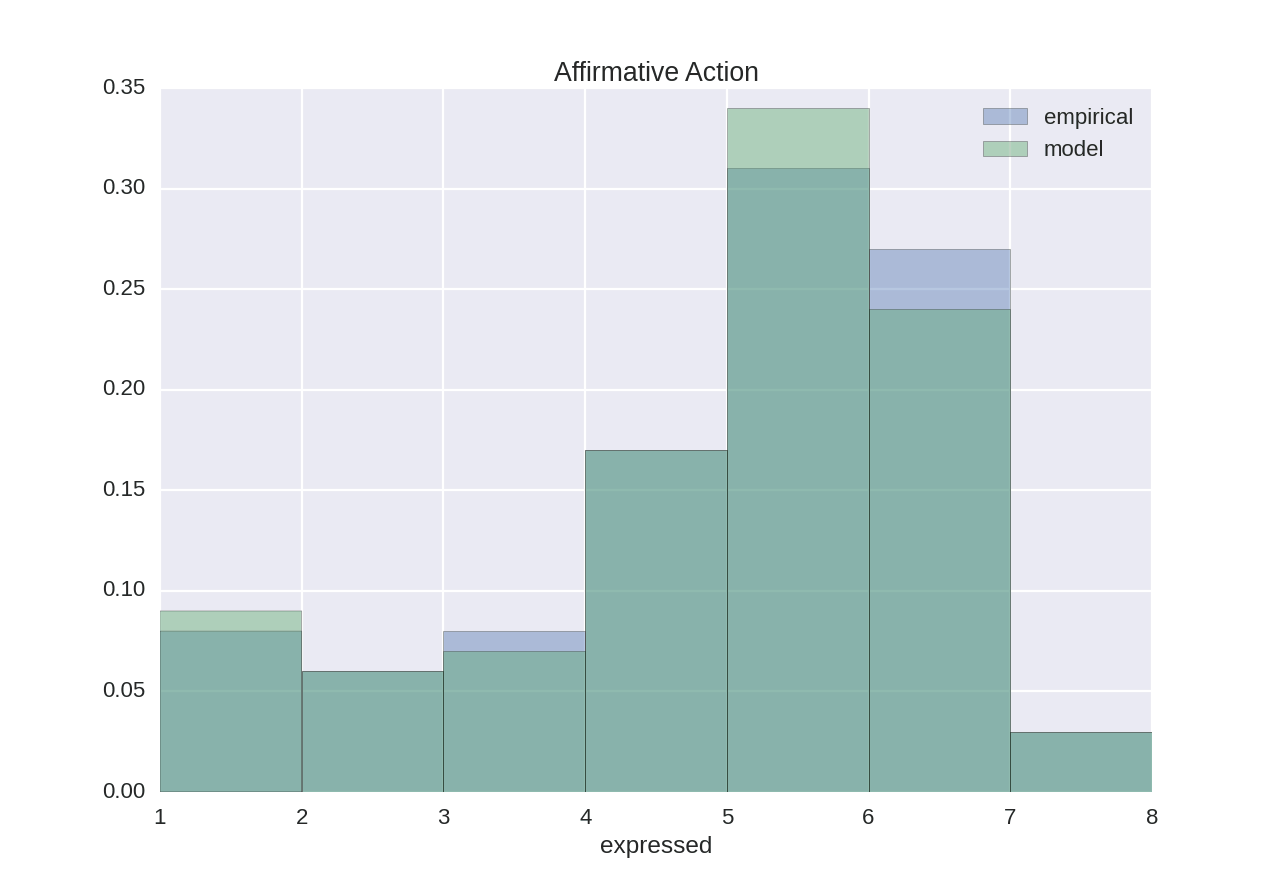}
\includegraphics[width=0.49\textwidth]{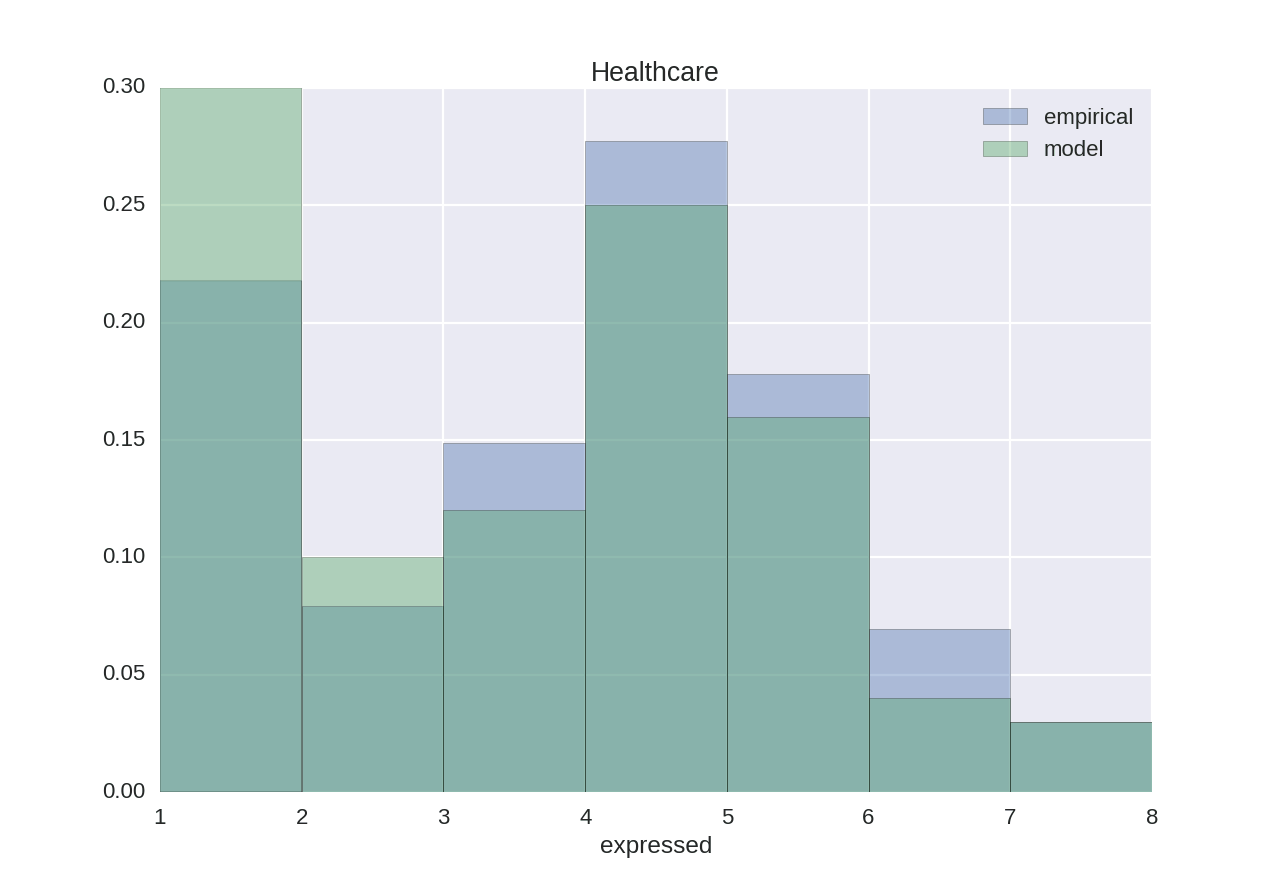}
\includegraphics[width=0.49\textwidth]{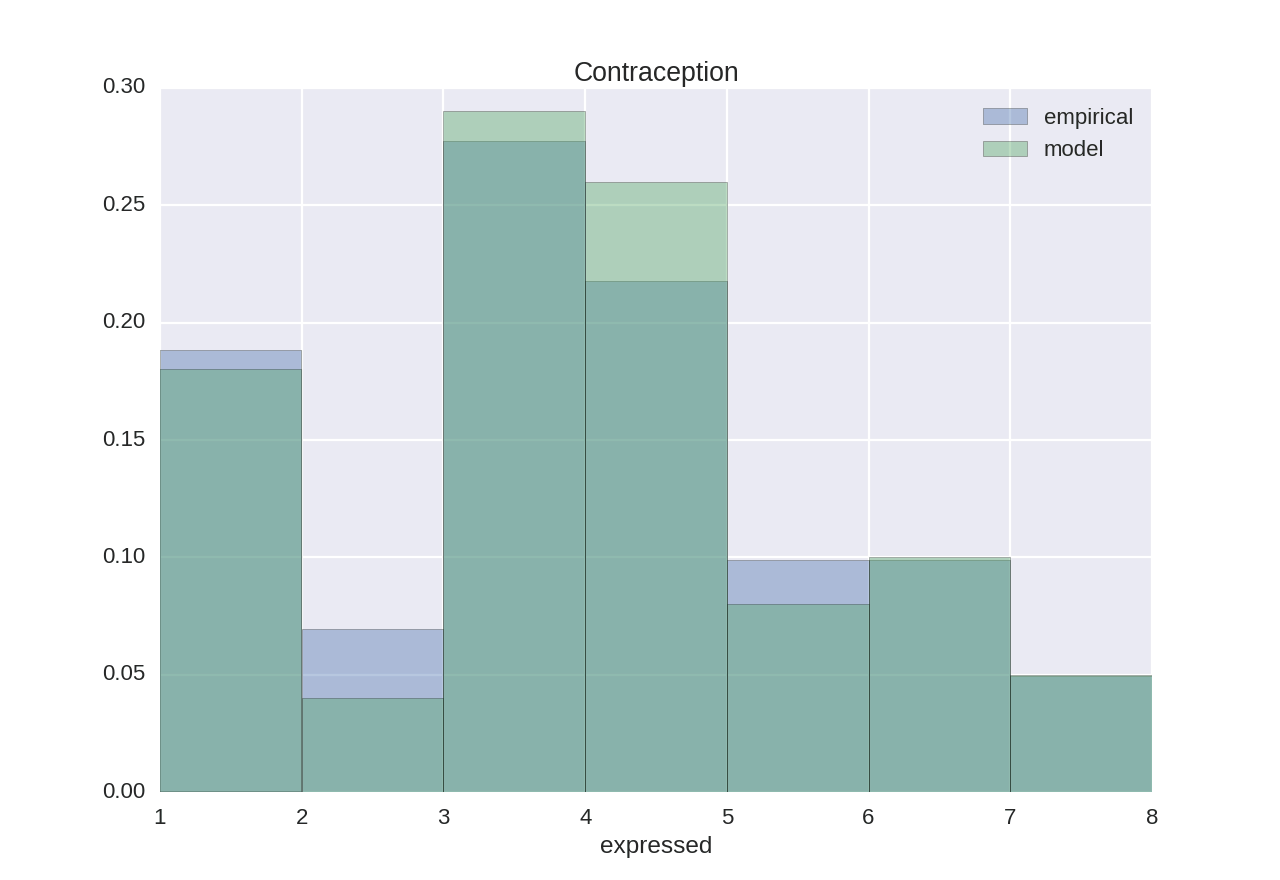}
\includegraphics[width=0.49\textwidth]{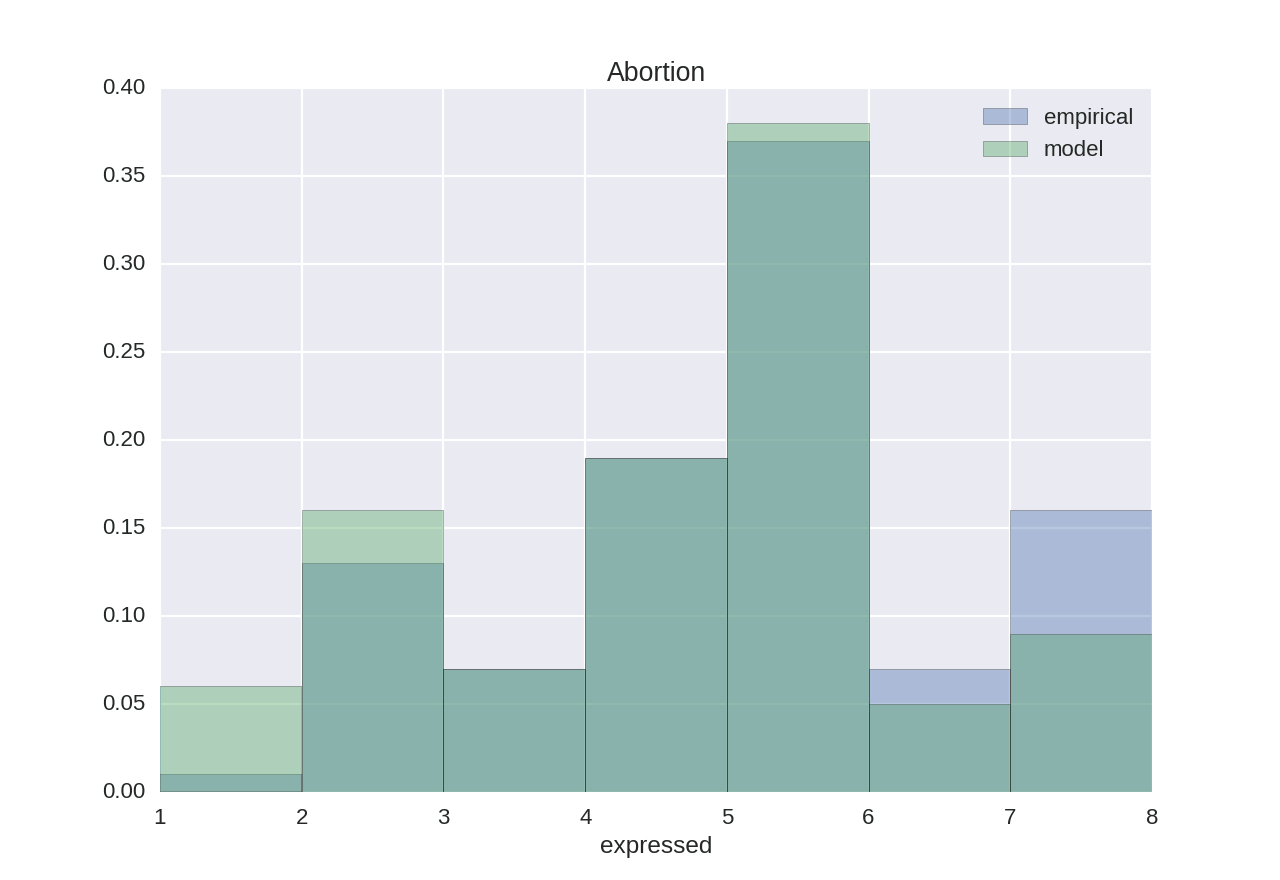}
\includegraphics[width=0.49\textwidth]{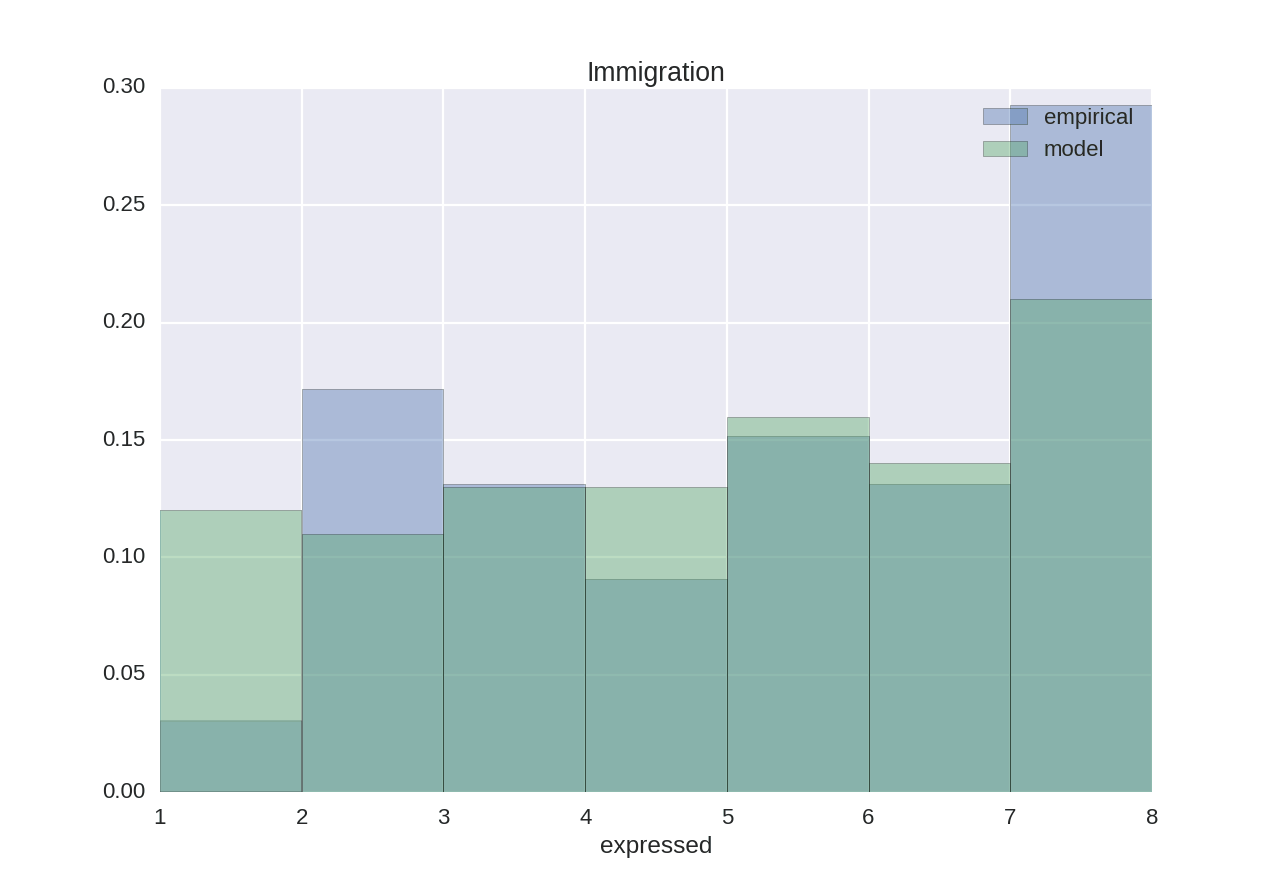}
\caption{The ISC model produces expressed opinion distributions that align with American's opinions on contemporary political issues ranging from gun control to healthcare to immigration. Data reproduced with permission from \cite{broockman2016}, parameters for each realization available on \href{https://github.com/psipeter/influence_susceptibility_conformity}{GitHub}}
\label{broockman}
\end{figure*}

As a proof-of-concept for strong diversity, I compare the expressed opinion distributions produced by the ISC model with a survey that assessed people's opinions on each of twelve issues in contemporary American politics \cite{broockman2016}. Each respondent was asked which of seven idealized positions, ranging from extremely liberal to extremely conservative statements about that issue, best described his or her belief, creating a seven-point opinion scale. Using several parameter-space exploration strategies, including an evolutionary algorithm and \verb!hyperopt!\cite{bergstra2015}, I found values for mean intolerance, susceptibility, and conformity that produced the distributions shown in Figure \ref{broockman}. These parameters, optimized to reduce the root-mean-square-error between the model distribution and Broockman's data over $n=4$ realizations, lie within the bounds of the values used in the above experiments (except for $\sigma_O=50$). The model captures distributions with a variety of different shapes, including: normal distributions around a centrist opinion (gun control) and a moderate opinion (affirmative action); centrist dominance with an extreme group (healthcare and contraception); and other strange shapes (abortion and immigration, with less accuracy). This result quantitatively demonstrates that real-world opinion distributions are within the output-space of the model.

\begin{figure*}[p]
\centering
\includegraphics[width=0.49\textwidth]{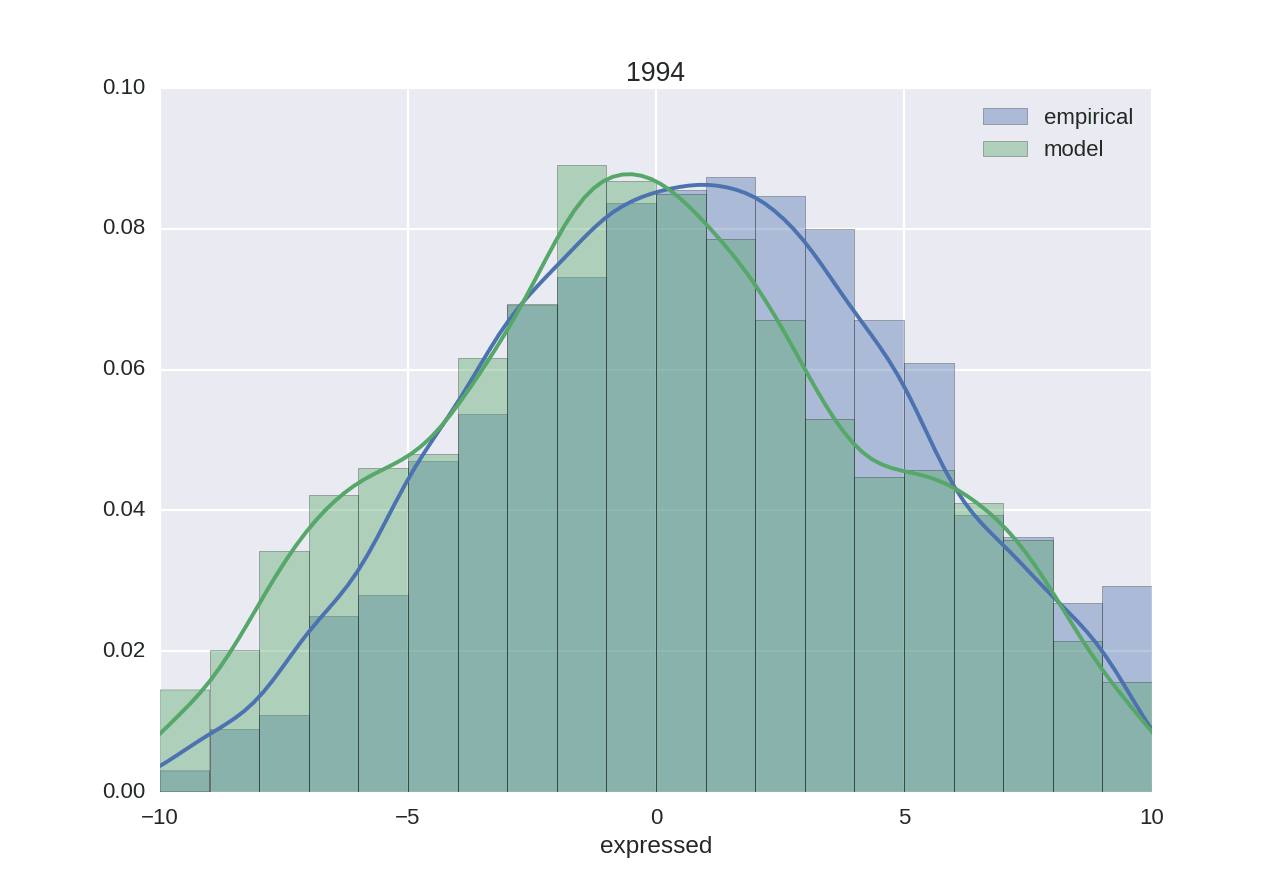}
\includegraphics[width=0.49\textwidth]{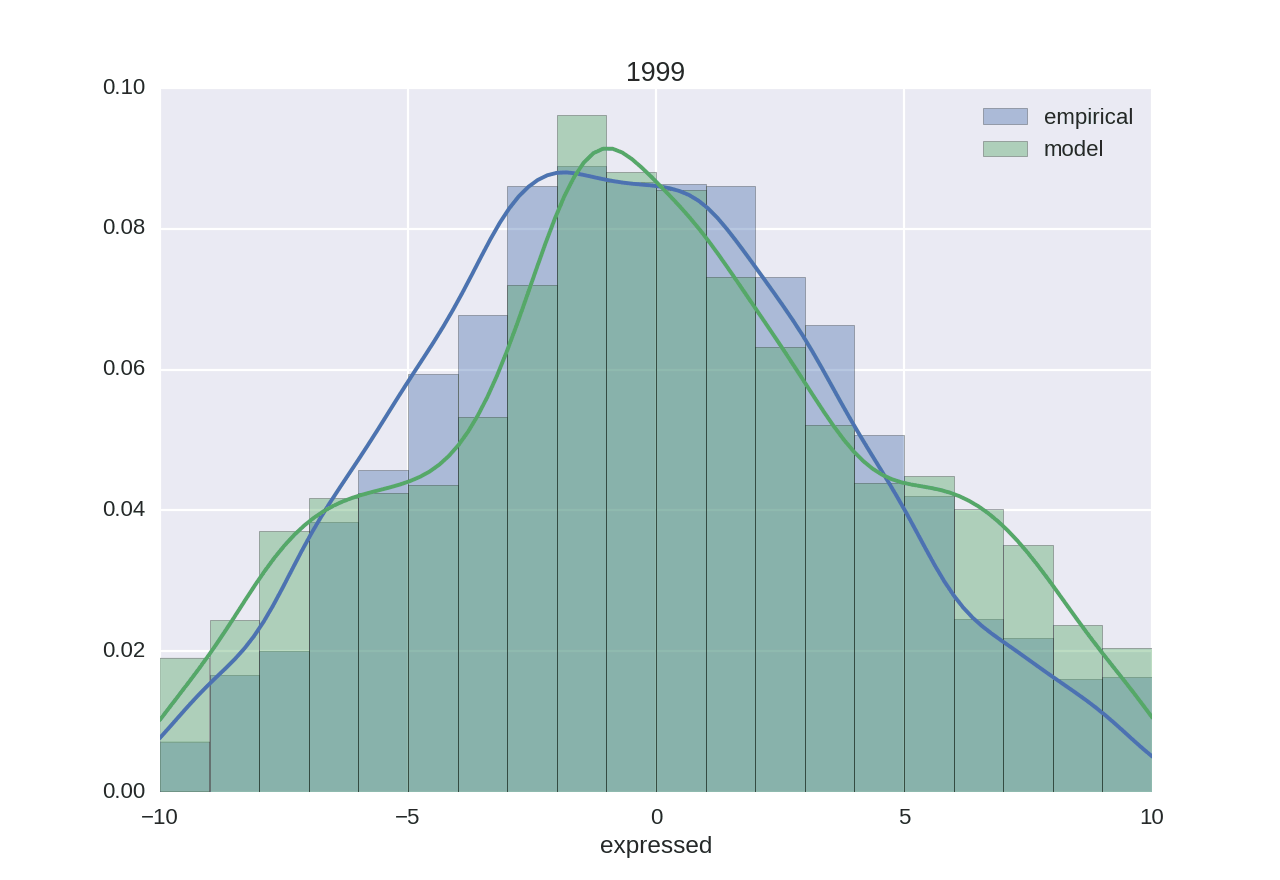}
\includegraphics[width=0.49\textwidth]{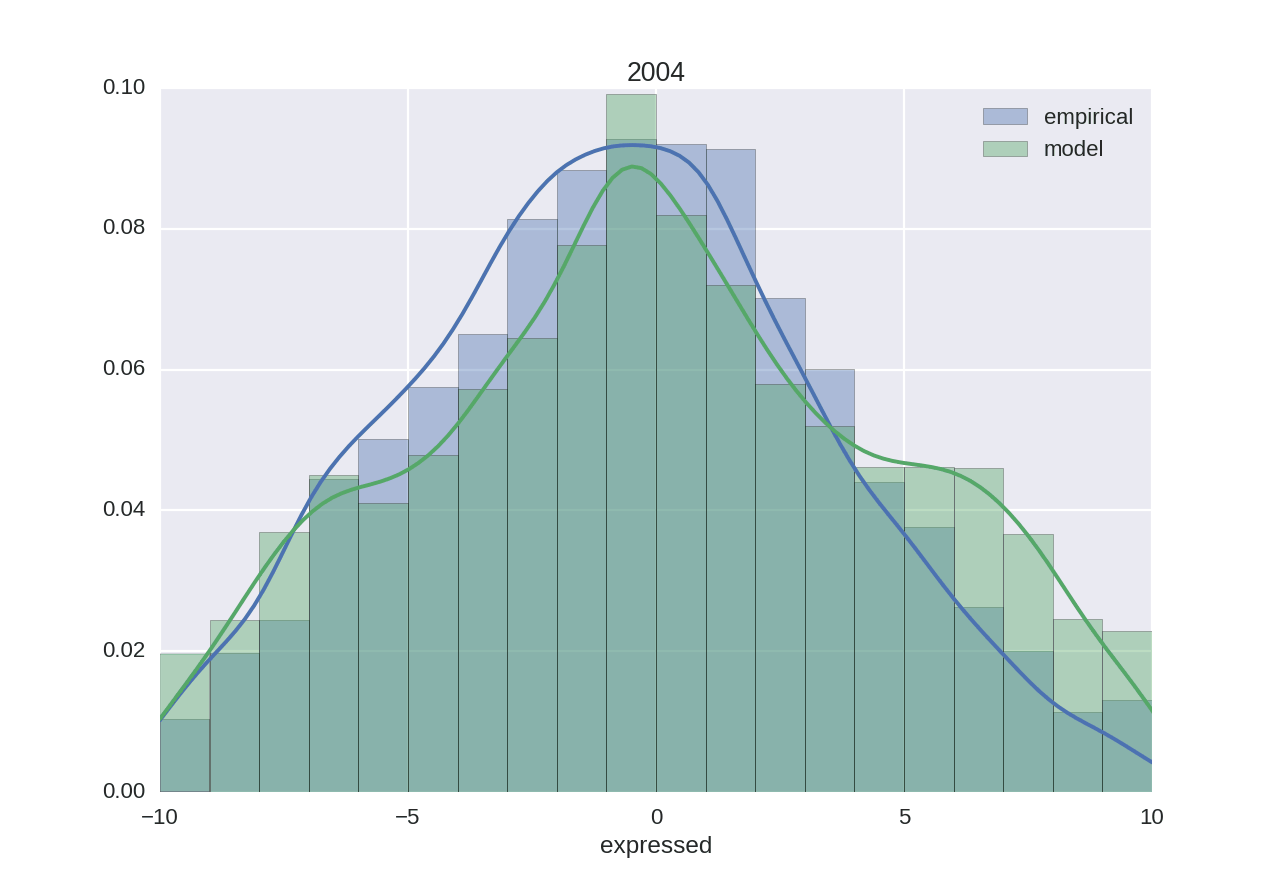}
\includegraphics[width=0.49\textwidth]{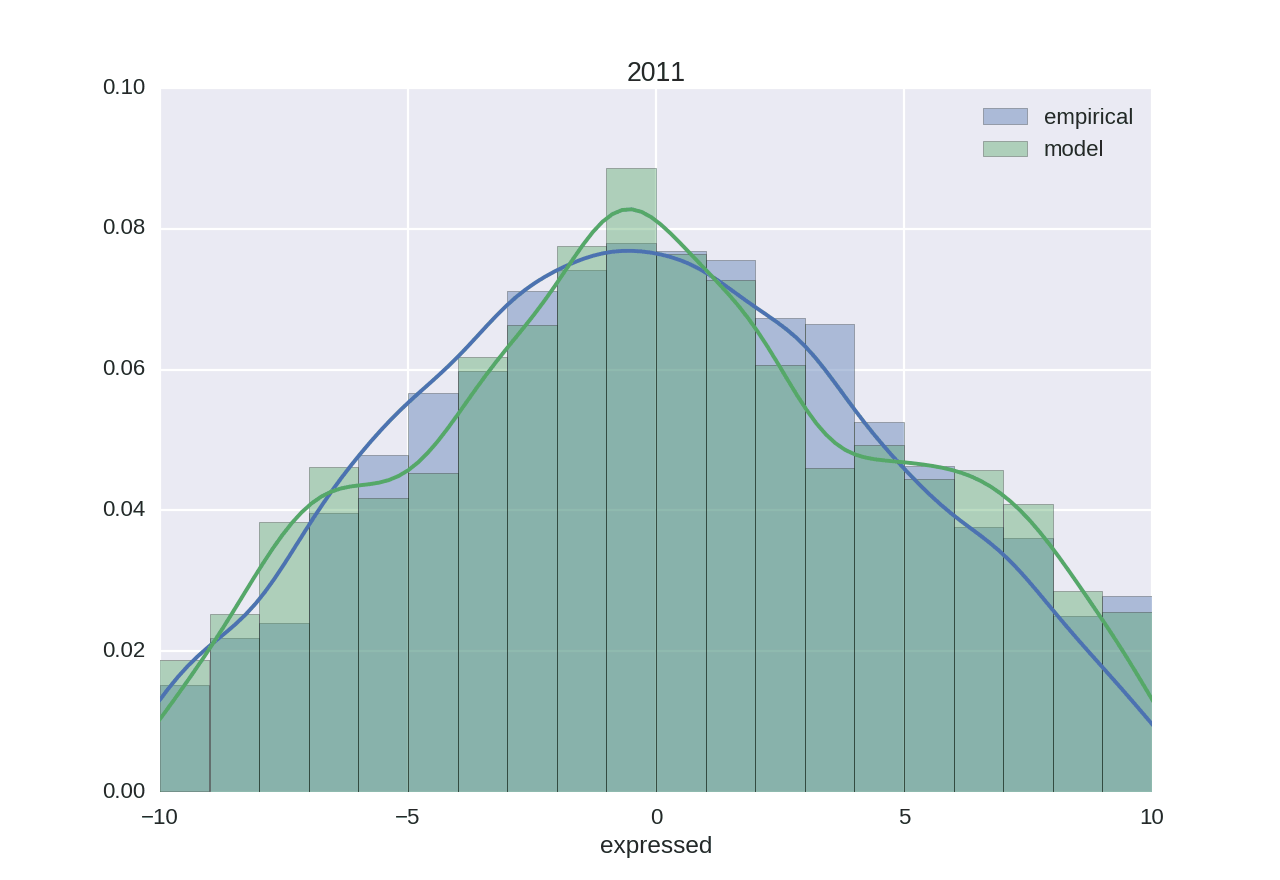}
\includegraphics[width=0.49\textwidth]{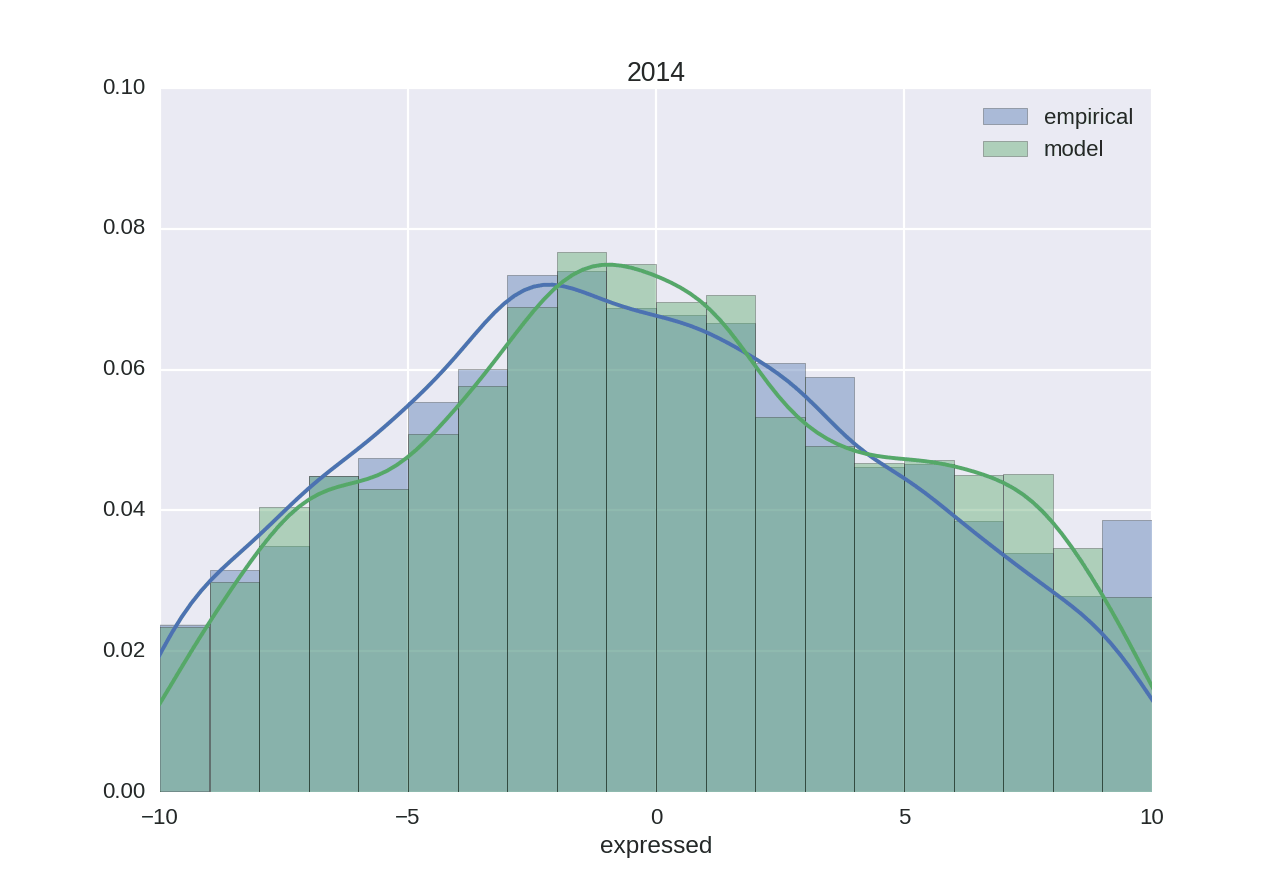}
\caption{The ISC model produces opinion dynamics that are consistent with the polarization of Americans' political opinions from 1994 to 2014. Blue and green lines are Gaussian kernel density estimates for the respective distributions. Data reproduced from the Pew Research Center \cite{pew2014}}
\label{pew}
\end{figure*}

I also compare the model's opinion dynamics with a large-$N$, multi-year survey of American's ideological consistency conducted by the Pew Research Center \cite{pew2014}. The survey consisted of ten questions assessing individuals' attitudes about current political issues such as ``[the] size and scope of government, the social safety net, immigration, homosexuality, business, the environment, foreign policy and racial discrimination,'' with each response coded  $-1$ (liberal), $+1$ (conservative) , or $0$ (don't know/refused). These values were summed for each individual, creating an ``ideological consistency'' scale ranging from $-10$ (liberal responses to every question) to $+10$ (conservative responses to every question). The study found that Americans have become increasingly polarized from 1994 to 2014: individuals who previously held mixed liberal and conservative positions on different issues are increasing partisan and ideologically uniform. As shown in Figure \ref{pew}, this trend manifests a spreading of the empirical distribution over time. Using the same parameter-space exploration tools as above, I found the ISC model produced similar patterns of polarization: a normal-like opinion distribution midway through the simulation gradually spread as extremists on both sides pulled centrists towards the periphery. It can also reproduce more subtle dynamics, such as the leftward shift of the kernel density estimate's central peak from 1994 to 1999, then back to the right as a sharper peak from 1999 to 2004. 



\section*{Discussion}
\addcontentsline{toc}{section}{Discussion}

\subsection*{How do social groups maintain a strong diversity of opinions?}
Psychological forces such as a commitment to strong opinions or a drive to distinctiveness may oppose homogenizing social influence and preserve strong diversity. When society contains agents with heterogeneous intolerance, susceptibility, and conformity, each individual is simultaneously pulled towards centrism and extremism. If these forces are balanced, a strong diversity of opinions emerges and remains stable through time. Figures \ref{diversity}, \ref{subcultures}, and \ref{halfreach} showed this diversity can take the form of (a) a centrist party diversified by influence from a few extremist, (b) two extremist parties with undecided agents on the borders, and (c) geographically-isolated opinion subcultures.

The maintenance of strong diversity is a novel result in opinion change and cultural diffusion models based on bounded confidence, which assume that agents influence one another only if their opinion similarity is above an interaction threshold \cite{macy2003, mark1998, hegselmann2002, deffuant2002, dandekar2013, salzarulo2006, jager2005}. Though this approximation of intolerance has proved a useful first step in understanding convergence vs. polarization, I argue that it is overly rigid: people do not classify each others' trustworthiness according to a binary scheme. The ISC model assumes that social influence changes continuously with intolerance, commitment, and context, and produces sustained, strong diversity under multiple psychological and network conditions. This result is intuitive, since societies do not converge to a single opinion or diverge to two polar opposites, and is also quantitatively plausible, as shown through empirical validation. Whenever possible, agent-based modelers should move away from psychologically and socially implausible assumptions and adopt empirically-motivated cognitive heuristics: doing so will solidify the model's foundations and, as exemplified by this study, produce more complex and realistic results.

Maintaining a diversity of opinions is important outside the modeling community. Indigenous cultures dissolve in the face of globalization as people substitute traditional languages and practices for the norms of modern society. Corporations fall prey to groupthink when individuals with original ideas choose not to voice them. Political and religions groups become polarized due to intolerance of dissimilar beliefs. To promote cultural and ideological diversity, leaders must recognize that social influence is not the only force that drives single-mindedness. They must recognize, not just conceptually but with the quantitative precision afforded by computational models, the role of psychological heterogeneity, personal commitment, and social context in destroying the valuable resource of diversity. 

\subsection*{Will subcultures of opinions survive in a well connected society?}
In the ISC model, communities of dissenters can survive among globalized centrism or extremist competition in two circumstances. In a tolerant or conformist society, the push towards centrism rapidly homogenizes most agents, but leaves a few intolerant or committed agents on the ideological periphery. When intolerant agents who hold opposing beliefs live together, they reject the opposite perspective so strongly that they become extreme despite centrist influence, as in Experiments 3 and 5. Polarization spreads outward, leaving cohesive extremist parties and undecided agents on the neighborhoods' borders. Densely clustered extremists resist moderate influence; it seems that neither a centrist majority nor an opposing extreme minority effectively moderates their speech or prevents their polarizing influence. If left undisturbed, these individuals will either settle into small conflicted communities or radicalize society. 

In the second scenario, agents' small social networks limit communication, producing a larger number of semi-isolated, cohesive, persistent communities. Communication is still possible between such communities, but must travel through bridging individuals whose influence is often overcome by the group consensus. These communities cannot coalesce when agents' social networks are large; social influence, when distributed over a large network, may cause either centrist convergence or extremist takeover. These results imply that (a) a lack of communication within society can encourage ideological splintering in the same way that geographic barriers facilitate speciation and genetic diversification, and (b) when advances in communication technology put isolated cultures into contact with the outside world, the inflow of globalized ideas can overwhelm the distinct features of their culture. In reality, extremism often emerges in locations with limited communication and access to external information. Although networking extremists with new individuals has the potential to spread radicalization, it also increases the probability that extremists will find a bridge to more moderate attitudes that, over time, persuades them to soften their beliefs, as occurred in Experiment 6. 

\subsection*{Does pluralistic ignorance affect societal opinion change?}
We assess others' opinions through their expressions and use that assessment to reevaluate our own beliefs. Experiment 2 showed that agents' desire to conform can lead others to mistakenly think that agreement exists in society, reverse the process of opinion polarization, and bring an intolerant society back to consensus. It also showed that when agents express opinions with the goal of appearing distinct, no observable consensus exists on any issue, and the false atmosphere of extremism causes societal polarization. Social norms like the desire to be distinct from the previous generation or to conform to the community's religious beliefs do have far reaching effects on opinion change at the societal level; any simulation which assumes agents have perfect information is missing an important aspect of social communication.

Pluralistic ignorance appeared in simulations with nonlinear opinion dynamics, spiking during the critical periods of change in that society's history, such as before stubborn extremist converted to centrism and when centrist experimented with moderate expressions. One interpretation of these results is that long-term history is relatively predictable when everyone communicates perfectly, but when social context encourages belief falsification, tensions between what is heard and what is felt build until they are suddenly released. This interpretation agrees with Kuran's work on the role of preference falsification in authoritarian revolutions \cite{kuran1989}, and was likely a contributing factor in the unexpected and rapid nature of the arab spring \cite{goodwin2011}.

\subsection*{Can the ISC model reproduce empirical opinion distributions and dynamics?}
All models should be treated with skepticism until they have been credibly validated with empirical data. The agreement between Broockman's data and the simulated opinion distributions shows that the model reproduces strong diversity. These political opinion distributions are sometimes far from normal, and may be non-symmetric or have few agents at the ideological center. Furthermore, the similarities between political polarization in the Pew dataset and in the simulation shows the model also captures certain features of opinion change, including short-term centrist fluctuations and long-term societal polarization. Overall, the validation experiments should be seen as an existence proof of plausible diversity and dynamics in the model, not as evidence of a calibrated simulation capable of precisely predicting opinion change.

\section*{Conclusion}
\addcontentsline{toc}{section}{Conclusion}

In this study, I examined the relationship between the psychosocial forces driving opinion change and the resulting distributions, dynamics, and topologies of opinions across society. This research extends previous studies in computational opinion dynamics by expanding the psychological depth of agents to include previously unstudied forces. Through a series of computational experiments, I showed that networks of heterogeneous agents will interact to produce (a) distributions of opinions that match political opinion data (b) opinion subcultures, and (c) trend-setting pluralistic ignorance. These results are significant advances in the study of macroscopic opinion change and suggest that modest increases in the complexity of agent models can produce opinion dynamics that align better with reality. 

Many extensions of the ISC model are possible. People actively promote their opinions at rallies or online, while others join organizations that enforce their beliefs through coercion and punishment. Introducing social mechanisms for these behaviors would permit the study of collective action problems and suggest more specific strategies that leaders could take to achieve desired patterns of opinion change. Another extension would allow for dynamic social networks. Though the social reach procedure captures important statistics of social networks, the people with whom we converse change constantly. Introducing dynamic networking, possibly in an expanded virtual environment, would permit a more complete study of how opinions change in a society dominated by social media. I would also like to compare opinion geography and pluralistic ignorance to empirical data.

I contend that empirically-accurate patterns of opinion change only emerge when agents act according to plausible rules, and that modelers must expand the depth of agents' social cognition to explain complex social phenomenon. This is best achieved by endowing agents with human-like cognitive architectures capable of affecting perception, memory, emotion, attention, and communication. Several opinion change models have already incorporated neurally-inspired mechanisms to great effect \cite{schroder2013,wolf2015}. Recent advances in neural engineering suggest that building agents with artificial brains may soon be possible \cite{eliasmith2012}. In future work, I plan to incorporate such artificial intelligences into social simulations.

\phantomsection
\bibliographystyle{unsrt}
\bibliography{bibliography}
\end{document}